\documentclass{ametsocV6.1_without_lineno}

\title{A Mechanism of Stochastic Synchronization in the Climate System: an Interpretation of the Boundary Current Synchronization as a Maxwell's Demon}

\authors{
    Yuki Yasuda\aff{a}\correspondingauthor{Yuki Yasuda, yasuda.yuki@scrc.iir.isct.ac.jp}
    and Tsubasa Kohyama\aff{b}
}

\affiliation{
    \aff{a}{Supercomputing Research Center, Institute of Integrated Research, Institute of Science Tokyo}\\
    \aff{b}{Department of Information Sciences, Ochanomizu University}
}

\abstract{This study has applied information thermodynamics to a bivariate linear stochastic differential equation (SDE) that describes a synchronization phenomenon of sea surface temperatures (SSTs) between the Gulf Stream and the Kuroshio Current, which is referred to as the boundary current synchronization (BCS). Information thermodynamics divides the entire system fluctuating with stochastic noise into subsystems and describes the interactions between these subsystems from the perspective of information transfer. The SDE coefficients have been estimated through regression analysis using observational and numerical simulation data. In the absence of stochastic noise, the solution of the estimated SDE shows that the SSTs relax toward zero without oscillation. The estimated SDE can be interpreted as a Maxwell's demon system, with the Gulf Stream playing the role of the ``Particle'' and the Kuroshio Current playing the role of the ``Demon.'' The Gulf Stream forces the SST of the Kuroshio Current to be in phase. By contrast, the Kuroshio Current maintains the phase by interfering with the relaxation of the Gulf Stream SST. In the framework of Maxwell's demon, the Gulf Stream is measured by the Kuroshio Current, whereas the Kuroshio Current performs feedback control on the Gulf Stream. When the Gulf Stream and the Kuroshio Current are coupled in an appropriate parameter regime, synchronization is realized with atmospheric and oceanic fluctuations as the driving source. This new mechanism, ``stochastic synchronization,'' suggests that such fluctuations can be converted into directional variations in a subsystem of the climate system through synchronization by utilizing information (i.e., information-to-energy conversion).}

\begin{document}

\maketitle

%
%
%
%
%
%

%

\section{Introduction} \label{sec:introduction}

Stochastic modeling is an effective approach for various climate phenomena \citep[e.g.,][]{Dijkstra2013Book, Franzke+2015WIREClimChange}. This approach is based on the time scale separation between slow and fast variations \citep{Hasselmann1976Tellus}. Typically, slow processes include decadal climate variations, while fast processes include weather disturbances. The scale separation suggests that the correlation between slow and fast processes is weak; thus, the fast processes can effectively be represented by stochastic noise. Stochastic modeling describes the integral (or aggregated) responses of slow climate variables driven by stochastic noise. A major approach in stochastic modeling is the use of idealized dynamical systems with few variables \citep[e.g.,][]{Dijkstra2013Book}. Although these models cannot describe phenomena in detail, they are expected to capture the essence of phenomena \citep[e.g.,][]{Benzi2010NPG}. In this study, we investigate the coupling of two western boundary currents using a stochastic dynamical system and propose a new mechanism for synchronization, which may be applicable to other climate phenomena.

Western boundary currents (WBCs) are strong upper flows located on the western side of ocean basins, playing an important role in the transport of heat in the climate system \citep{Kwon+10JCLI}. Two major examples of WBCs are the Gulf Stream in the North Atlantic and the Kuroshio Current in the North Pacific. These WBCs are primarily driven by atmospheric wind stress \citep{Stommel48} and modulated through interactions with the atmosphere \citep{Qiu+14JCLI,Ma+16Nature}. The heat transport by the WBCs influences the entire atmosphere, including storm tracks and annular modes \citep{Hoskins+Valdes90JAS,Minobe+08Nature,Ogawa+12GRL,Omrani+19SciRep}. Furthermore, the WBCs sometimes contribute to explosive cyclogenesis, making them also important for daily weather \citep{Sanders86MWR,Yoshida+Minobe17JCLI}.

The WBCs, namely the Gulf Stream and the Kuroshio Current, exhibit interannual to decadal variability \citep{Qiu+Chen05JPO,Kelly+10JCLI}. Decadal variability due to the interaction of either the Gulf Stream or the Kuroshio Current with the atmosphere has been studied \citep{Latif+Barnett94Science,Gallego+Cessi00ClimDyn,Hogg+06JCLI,Kwon+10JCLI,Qiu+14JCLI}. Recently, the importance of the coexistence of both the Gulf Stream and the Kuroshio Current has been pointed out for ocean heat content \citep{Kelly+Dong04EarthClim} and atmospheric annular modes \citep{Omrani+19SciRep}. However, the roles of both ocean currents in the climate system are not yet fully understood.

\citet{Kohyama+21Science} discovered the boundary current synchronization (BCS) by using observational data and numerical experiments with coupled atmosphere-ocean models. The sea surface temperatures (SSTs) of the Gulf Stream and the Kuroshio Current show positive correlations on interannual to decadal time scales. An increase (decrease) in the SSTs is accompanied by a northward (southward) shift of the WBCs as well as a northward (southward) shift of the atmospheric baroclinic jet. The BCS is considered to arise from the interaction between the Gulf Stream and the Kuroshio Current via the atmosphere (i.e., the inter-basin coupling) rather than a passive response to atmospheric forcings. In fact, the BCS was not reproduced in low-resolution numerical experiments that could not resolve oceanic mesoscale eddies. These results suggest that the BCS occurs through the transfer of information on WBC variability through the atmosphere. However, the roles of the Gulf Stream and the Kuroshio Current in the BCS have not been well investigated, and further research is required.

Over the past decade, the fusion of information theory and non-equilibrium physics has led to the development of a new theoretical framework called ``information thermodynamics'' \citep{Parrondo+15NatPhys,Peliti+Pigolotti21Book,Shiraishi23Book}. Information thermodynamics divides nonlinear systems fluctuating with stochastic noise into subsystems and describes the interactions between these subsystems from the perspective of information transfer. Information transfer can be expressed as information flow \citep{Allahverdyan+09JStatMech} and is incorporated into the second law of thermodynamics \citep{Horowitz+Esposito14PhysRevX}. Applying this extended second law of thermodynamics (i.e., the second law of information thermodynamics), we can elucidate the dynamics of subsystems and the asymmetry between them. This new theoretical framework is applicable to stochastic systems that include deterministic forcing and dissipation. These points suggest that information thermodynamics is suitable for studying the climate system, which is a nonlinear system with multiple degrees of freedom driven by stochastic noise and deterministic forcings. However, since information thermodynamics is an emerging field, it has scarcely been used in atmospheric and oceanic sciences.

This study applies information thermodynamics to a dynamical system for the BCS and proposes that the coupled system of the Gulf Stream and the Kuroshio Current can be interpreted as a Maxwell's demon. Section \ref{sec:maxwell-demon} provides a review of Maxwell's demon. Section \ref{sec:dynamical-system} introduces the bivariate linear dynamical system that describes the time evolution of the regional-mean SSTs of the Gulf Stream and the Kuroshio Current. Section \ref{sec:analysis} applies information thermodynamics to this system and discusses the roles of the Gulf Stream and the Kuroshio Current. Section \ref{sec:conclusions} presents the conclusions.

\section{Review of Maxwell's demon} \label{sec:maxwell-demon}

The resolution of Maxwell's demon is a landmark in information thermodynamics \citep{Parrondo+15NatPhys}. Originally, Maxwell argued that if an intelligent being (a demon) had information about the velocities of particles, it could separate fast, hot particles from slow, cold particles without exerting any work on the system, thus appearing to violate the second law of thermodynamics. Szilard then developed a simple model, the Szilard engine, which captures the essential point of Maxwell's demon: information can be converted to work. Nowadays, this information-to-energy conversion has been realized experimentally \citep{Toyabe+2010NatPhys, Koski+14PNAS}. This type of system is called an information engine or loosely Maxwell's demon \citep[e.g.,][]{Horowitz+Esposito14PhysRevX, Ito+Sagawa15NatComm}. Here, we refer to the entire system of information conversion as the Maxwell's demon system and review the Szilard engine as such a system, which consists of a demon (i.e., a controller) and a subsystem operated by the demon. See further details in reviews and textbooks \citep{Sagawa+Ueda13InBook, Parrondo+15NatPhys, Peliti+Pigolotti21Book, Shiraishi23Book}.

\subsection{Setup of the Szilard engine} \label{subsec:szilard-engine}

\begin{figure}[t]
    \centering
    \noindent\includegraphics[width=14cm]{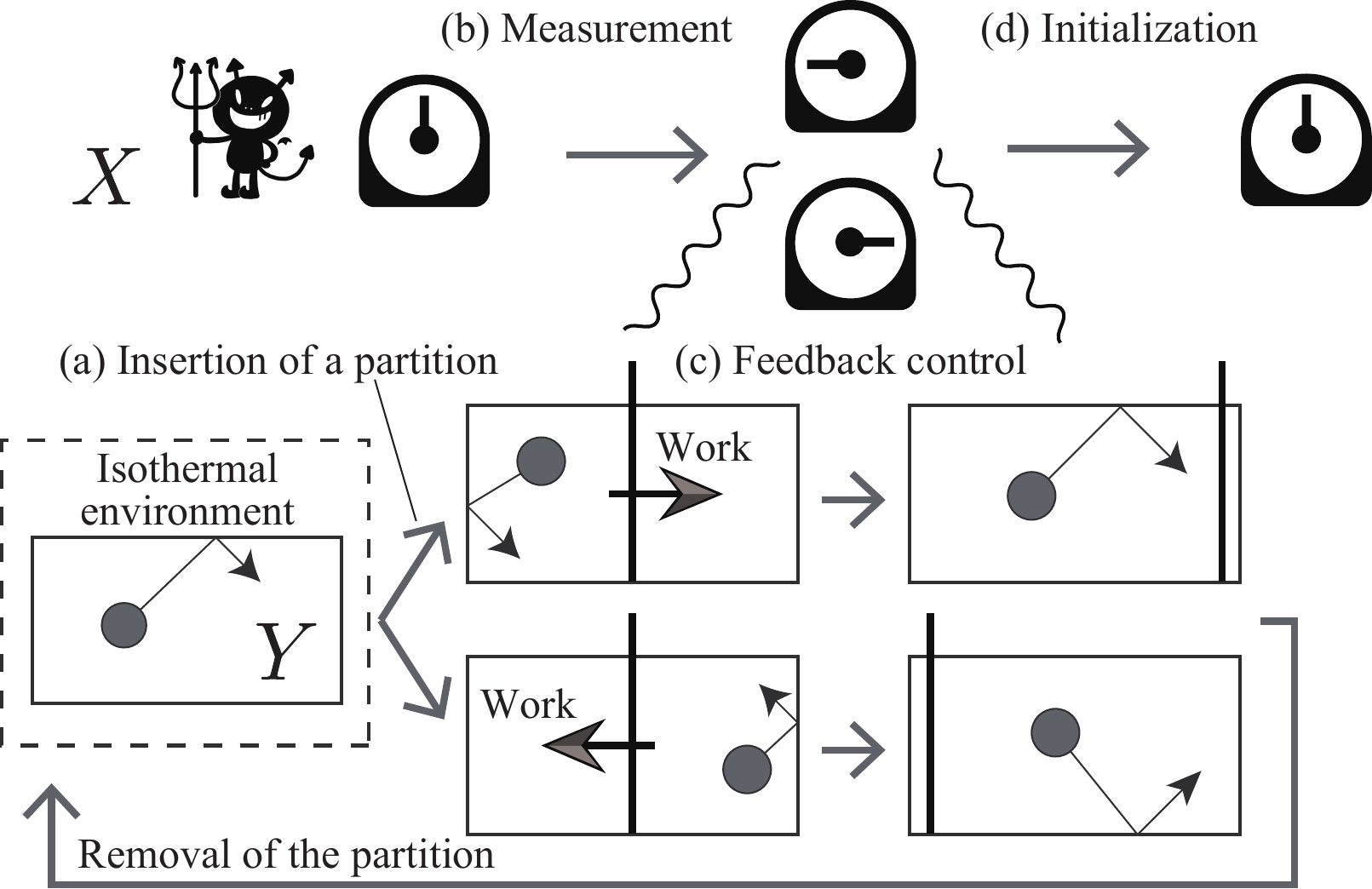}
    \caption{Schematic of the Szilard engine \citep[e.g.,][]{Parrondo+15NatPhys}. A box contains a single particle and is surrounded by an isothermal environment, where heat is transferred through the walls of the box. (a) A partition is inserted in the middle of the box. (b) The demon measures the position of the particle and records the result in its memory. (c) Based on the measurement outcome, the demon performs feedback control. Specifically, when the particle is on the left (resp. right), the demon moves the partition to the right (resp. left). The particle system is isothermally expanded and returns to the initial state. This process extracts a positive work from the isothermal environment. (d) The demon's memory is initialized, requiring external work (i.e., a negative work).}
    \label{fig:szilard-demon-particle}
\end{figure}

Let us consider an isothermal cycle with a single particle. Figure \ref{fig:szilard-demon-particle} is a schematic of this experiment, where the memory (i.e., the demon) in the upper half is not considered for the moment. First, a partition is inserted in the middle of the box, separating the box into halves (Fig. \ref{fig:szilard-demon-particle}a). The particle moves at random, so the probability of the particle being on the left (or right) side equals $1/2$. Second, the position of the particle is measured (Fig. \ref{fig:szilard-demon-particle}b). Third, based on the measurement outcome, when the particle is on the left (resp. right), the partition is moved quasi-statically to the right (resp. left). (Fig. \ref{fig:szilard-demon-particle}c). Such a measurement-based operation is called feedback control or simply, control. The energy source for this expansion is the surrounding thermal fluctuations, or equivalently, heat transferred through the box. Therefore, the Szilard engine appears to contradict the second law of thermodynamics and to be a perpetual motion machine of the second kind that extracts work via isothermal cycles.

To resolve the paradox, it is necessary to consider the exchange of information quantities and the existence of memory \citep{Sagawa+Ueda13InBook}. The memory, also called the demon, is a subsystem that performs measurement and feedback control. The other subsystem is the particle within the box and is referred to as the particle. We denote the states of the demon and the particle by $X$ and $Y$ and their realizations by lowercase letters $x$ and $y$, respectively. The variables $X$ and $Y$ are discrete, reflecting the particle position being either on the left or right. We first derive the second law of information thermodynamics and then consider the Szilard engine again.

\subsection{The second law of information thermodynamics} \label{subsec:second-law-}

We define Shannon entropy and mutual information \citep[e.g.,][]{Cover+Thomas05Book} to derive the second law of information thermodynamics.
\begin{align}
    S(X) & :=\sum_x p(x)[-\ln p(x)], \label{eq:entropy-S-X} \\
    S(Y) & :=\sum_y p(y)[-\ln p(y)], \label{eq:entropy-S-Y} \\
    M(X, Y) & :=\sum_{x, y} p(x, y) \ln \left[\frac{p(x, y)}{p(x) p(y)}\right]. \label{eq:mutual-info-X-Y}
\end{align}
Shannon entropies $S(X)$ and $S(Y)$ are indicators of uncertainty about the subsystems $X$ and $Y$, respectively. Mutual information $M(X,Y)$ is a non-negative indicator of the correlation between $X$ and $Y$, where $M(X,Y)$ is zero only when $X$ and $Y$ are statistically independent.

Mutual information characterizes the non-additivity of Shannon entropy \citep[e.g.,][]{Cover+Thomas05Book}. For the joint Shannon entropy $S(X,Y)$, the following holds:
\begin{align}
    S(X, Y) & :=\sum_{x, y} p(x, y)[-\ln p(x, y)], \label{eq:non-additivity-S-XY-1} \\
    & =S(X)+S(Y)-M(X, Y). \label{eq:non-additivity-S-XY-2}
\end{align}
Adding the uncertainties for the subsystems, $S(X)+S(Y)$, leads to an overestimation and is generally not equal to the uncertainty for the entire system $S(X,Y)$. It is necessary to subtract $M(X,Y)$, which is non-negative and represents the magnitude of correlation between $X$ and $Y$.

The second law of information thermodynamics is derived using the non-additivity of Shannon entropy \citep{Sagawa+Ueda13NewJPhys}. We consider the evolution from time $t$ to $t+\Delta t$ and assume the second law of thermodynamics for the entire system \citep{Seifert05PRL,Seifert12RepProgPhys}:
\begin{equation}
    \Delta S\left(X_t, Y_t\right)+\frac{\Delta Q_t}{T} \ge 0, \label{eq:second-law-total-XY}
\end{equation}
where changes are denoted by $\Delta$, and time dependence is denoted by a subscript $t$. For example, $\Delta S(X_t, Y_t) := S(X_{t+\Delta t}, Y_{t + \Delta t}) - S(X_t, Y_t)$. The symbol $\Delta Q_t$ represents the amount of heat dissipated from the joint system $(X,Y)$ to the surrounding environment, which corresponds to the entropy change of the environment. Equation (\ref{eq:second-law-total-XY}) expresses that the total entropy change, including the environment, is non-negative.

The second law of information thermodynamics for the subsystem $X$ is derived from Eq. (\ref{eq:second-law-total-XY}) by assuming that only $X$ changes in time. Using the entropy decomposition [Eq. (\ref{eq:non-additivity-S-XY-2})], the following is derived:
\begin{equation}
    \Delta S\left(X_t, Y_t\right)+\frac{\Delta Q_t}{T} =\Delta S\left(X_t\right)+\underbrace{\Delta S\left(Y_t\right)}_{=0}-\Delta M\left(X_t, Y_t\right)+\frac{\Delta Q_t}{T} \ge 0, \label{eq:second-law-X1}
\end{equation}
where $\Delta S(Y) = 0$ because $Y_t$ does not change in time. This assumption also indicates that the entropy change of the environment $\Delta Q_t/T$ is due to the change in $X_t$. We finally obtain the following inequality:
\begin{equation}
    \Delta \sigma\left(X_t\right) \ge \Delta M\left(X_t, Y_t\right), \label{eq:second-law-X2}
\end{equation}
where
\begin{align}
    \Delta \sigma\left(X_t\right) &:= \Delta S\left(X_t\right)+\frac{\Delta Q_t(X_t)}{T}, \label{eq:total-entropy-X}\\
    \Delta M\left(X_t, Y_t\right) &= M(X_{t+\Delta t}, Y_{t}) - M(X_{t}, Y_{t}).
\end{align}
The total entropy change by $X_t$ is denoted as $\Delta \sigma(X_t)$ and is lower bounded by the change in mutual information $\Delta M\left(X_t, Y_t\right)$. Since $Y_t$ does not change, $\Delta M\left(X_t, Y_t\right)$ is attributed only to the change in $X_t$. The second law of information thermodynamics [Eq. (\ref{eq:second-law-X2})] indicates that $\Delta \sigma(X_t)$ can be negative when $\Delta M\left(X_t, Y_t\right) < 0$. Thus, the entropy of the subsystem $X$ can be decreased by utilizing the change in mutual information without heat transfer. This conclusion is consistent with thermodynamics because this discussion started from the second law of thermodynamics for the entire system [Eq. (\ref{eq:second-law-total-XY})]. Note that $\Delta \sigma(Y_t) \ge M(X_t, Y_{t+\Delta t}) - M(X_{t}, Y_{t})$ is derived when only $Y$ changes in time.

\subsection{Explanation of the Szilard engine} \label{subsec:resolution-maxwell-demon}

The second law of information thermodynamics is used to explain the Szilard engine. Time $t$ is not explicitly stated if there is no confusion. In the initial state, the states of the demon $X$ and the particle $Y$ are uncorrelated, that is, $M(X,Y)$ is zero. First, the demon inserts a partition (Fig. \ref{fig:szilard-demon-particle}a). This causes the particle $Y$ to be either on the right or the left. The mutual information $M(X,Y)$ remains zero because the demon's internal state (i.e., the memory $X$) has not changed.

Next, the demon measures the position of the particle and changes its memory $X$ to record the measurement outcome, which makes $X$ either right or left (Fig. \ref{fig:szilard-demon-particle}b). The states of the demon and the particle become correlated, and the mutual information increases, $\Delta M(X,Y) > 0$. This increase in correlation is due to the change in the demon $X$. According to the second law of information thermodynamics, the increase in correlation is accompanied by a positive entropy change, $\Delta \sigma(X) \ge \Delta M(X,Y)$ ($>0$) [Eq. (\ref{eq:second-law-X2})].

Based on the memory value $X$, the demon quasi-statically moves the partition to extract work (Fig. \ref{fig:szilard-demon-particle}c). Regardless of the memory value, the particle $Y$ returns to the initial state after this feedback control, and the correlation $M(X,Y)$ becomes zero again. This decrease in correlation is due to the change in the particle $Y$. According to the second law of information thermodynamics, the decrease in correlation can be accompanied by a negative entropy change, $\Delta \sigma(Y) \ge \Delta M(X,Y)$ ($< 0$). This negative $\Delta \sigma(Y)$ corresponds to the positive work extracted by the isothermal expansion.

Finally, work is performed from the outside to initialize the demon's memory $X$ (Fig. \ref{fig:szilard-demon-particle}d). There is a trade-off for the memory between initialization and measurement, and it is more appropriate to consider both together \citep{Sagawa+Ueda09PRL,Sagawa+Ueda13NewJPhys,Sagawa19InBook}. The positive work extracted by the isothermal expansion and the negative work required for the initialization (and measurement) cancel each other out. No net positive work can be extracted using the Szilard engine. Therefore, the Szilard engine is not a perpetual motion machine of the second kind.

Information thermodynamics clarifies that the positive work extracted during the control process is due to the change in mutual information; that is, the total entropy of a subsystem can decrease by consuming mutual information. In this way, information thermodynamics allows for the analysis of subsystems from the perspective of information.

\subsection{Autonomous Maxwell's demon systems} \label{subsec:autonomous-maxwell-demon}

Maxwell's demon systems are also realized in an autonomous manner \citep{Horowitz+Esposito14PhysRevX,Loos+Klapp20NewJPhys}. The time evolution of an autonomous system can be expressed by stochastic differential equations without external interference, such as human manipulation. In such a system, one cannot isolate individual steps, such as measurement and control, since all steps can occur simultaneously and continuously over time. This is a major difference between the Szilard engine and autonomous processes. Thus, the distinction between the ``Demon'' and the ``Particle'' appears unclear.

The determination of which subsystem is the ``Demon'' or the ``Particle'' is based on the direction of information flow \citep{Allahverdyan+09JStatMech,Horowitz+Esposito14PhysRevX,Loos+Klapp20NewJPhys}.
\begin{align}
    & \dot{I}_{X \leftarrow Y}:=\frac{M\left(X_{t+{\rm d} t}, Y_t\right)-M\left(X_t, Y_t\right)}{{\rm d} t}, \label{eq:def-I-Y-to-X} \\
    & \dot{I}_{Y \leftarrow X}:=\frac{M\left(X_t, Y_{t+{\rm d} t}\right)-M\left(X_t, Y_t\right)}{{\rm d} t}, \label{eq:def-I-X-to-Y} \\
    & \frac{{\rm d}}{{\rm d} t}M\left(X_t, Y_t\right)=\dot{I}_{X \leftarrow Y}+\dot{I}_{Y \leftarrow X}. \label{eq:conservation-I-X-Y}
\end{align}
The information flow $\dot{I}_{X \leftarrow Y}$ from $Y$ to $X$ is the change in mutual information $M(X_t, Y_t)$ due to the change in $X$, whereas $\dot{I}_{Y \leftarrow X}$ is due to the change in $Y$. The sum of both information flows is equal to the total rate of change of mutual information [Eq. (\ref{eq:conservation-I-X-Y})]. Information flow is regarded as a quantity that represents the flow of information between subsystems.

For autonomous systems, the second law of information thermodynamics becomes as follows \citep{Horowitz+Esposito14PhysRevX,Loos+Klapp20NewJPhys}:
\begin{align}
    \dot{\sigma}(X) \ge \dot{I}_{X \leftarrow Y} \label{eq:second-law-info-autonomousX}, \\
    \dot{\sigma}(Y) \ge \dot{I}_{Y \leftarrow X} \label{eq:second-law-info-autonomousY}.
\end{align}
The rate of change of total entropy is lower bounded by the information flow. Hereafter, the rate of change of total entropy is referred to as the production rate of total entropy or simply the entropy production rate. Equations (\ref{eq:second-law-info-autonomousX}) and (\ref{eq:second-law-info-autonomousY}) are derived for general autonomous systems (Appendix A).

We finally summarize the distinction between a ``Demon'' and a ``Particle.'' As in Section \ref{sec:maxwell-demon}\ref{subsec:resolution-maxwell-demon}, measurement is a change that increases correlation due to an evolution of the ``Demon'', which is accompanied by a positive entropy production. On the other hand, control is a change that decreases correlation due to an evolution of the ``Particle'', which is accompanied by a negative entropy production. Therefore, in an autonomous system, the ``Demon'' (resp. ``Particle'') is the subsystem showing a positive (resp. negative) information flow and entropy production rate \citep{Horowitz+Esposito14PhysRevX,Loos+Klapp20NewJPhys}. In our study, a system consisting of a ``Demon'' and a ``Particle'' has been referred to as the Maxwell's demon system. Maxwell's demon systems are no longer thought experiments but have been experimentally realized in both non-autonomous and autonomous settings \citep{Toyabe+2010NatPhys,Koski+2015PRL}. We propose that such systems can also be realized in the atmosphere and ocean, that is, the coupled system of the Gulf Stream and the Kuroshio Current can be interpreted as a Maxwell's demon system.

\section{Bivariate linear dynamical system for BCS} \label{sec:dynamical-system}

\subsection{Governing equations} \label{subsec:equations-dynamical-system}

This section introduces a bivariate dynamical system describing BCS, which is hereafter referred to as the dynamical system. This system consists of the regional-mean SST anomalies of the Gulf Stream and the Kuroshio Current, $T_G$ and $T_K$, respectively. The subscripts $G$ and $K$ denote the Gulf Stream and the Kuroshio Current, respectively. Following \citet{Kohyama+21Science}, $T_G$ is the spatially averaged SST in the Gulf Stream region ($35^{\circ}{\rm N}$--$45^{\circ}{\rm N}$, $80^{\circ}{\rm W}$--$50^{\circ}{\rm W}$), and $T_K$ is the spatially averaged SST in the Kuroshio Current region ($35^{\circ}{\rm N}$--$45^{\circ}{\rm N}$, $140^{\circ}{\rm E}$--$170^{\circ}{\rm E}$). Both $T_G$ and $T_K$ are standardized (i.e., non-dimensionalized) after removing the climatology and the linear trend. 

The variables $T_G$ and $T_K$ are assumed to obey the following equations:
\begin{align}
    \frac{{\rm d} T_G}{{\rm d}t} &= -r_G T_G + c_{G \leftarrow K} T_K + \xi_G, \label{eq:dynamical-sys-TG} \\
    \frac{{\rm d} T_K}{{\rm d}t} &= -r_K T_K + c_{K \leftarrow G} T_G + \xi_K, \label{eq:dynamical-sys-TK} 
\end{align}
where $r_G$, $c_{G \leftarrow K}$, $r_K$, and $c_{K \leftarrow G}$ are real constants. The unit of time is set to $1$ month. The mutually independent white Gaussian noises $\xi_G$ and $\xi_K$ satisfy
\begin{align}
    \langle \xi_G(t) \xi_G(s)\rangle &= 2D_G \delta(t-s), \label{eq:auto-corr-xi_G}\\
    \langle \xi_K(t) \xi_K(s)\rangle &= 2D_K \delta(t-s), \label{eq:auto-corr-xi_K}\\
    \langle \xi_K(t) \xi_G(s)\rangle &= 0, \label{eq:cross-corr-xi_G-xi_K}
\end{align}
where $D_G$ and $D_K$ are positive constants, $\delta(t-s)$ is the Dirac delta function, $t-s$ is the time difference, and $\langle \cdot \rangle$ represents the ensemble average over the noise realizations.

This type of a system that consists only of oceanic variables (e.g., SSTs) can be derived from a more complicated setting that include both oceanic and atmospheric variables. Indeed, \citet{Gallego+Cessi01JClim} started with an atmosphere-ocean coupling model and then derived idealized equations that consist only of Atlantic and Pacific variables by representing the atmospheric effects in terms of the oceanic variables. Our dynamical system is interpreted as a simplified model derived through these procedures. The cross terms, $c_{G \leftarrow K} T_K$ and $c_{K \leftarrow G} T_G$, may include inter-basin couplings via the atmosphere.

\subsection{Regression analysis} \label{subsec:regression-analysis}

The coefficients of the dynamical system were estimated by the regression analysis using two sets of SST time series, both of which are used in \citet{Kohyama+21Science}. One is the observational data, Optimum Interpolation SST \citep[OISST,][]{Reynolds+07JCLI}. The other is the numerical output from a coupled atmosphere-ocean model, GFDL-CM4C192 \citep{Held+19JAMES,Eyring+16GMD}. Monthly mean values were used for both datasets. The OISST data range from December 1981 to September 2018, whereas the GFDL-CM4C192 data range from January 1950 to December 2014. Using the longer-period GFDL-CM4C192 data, we expect to enhance the reliability of the analysis results. All of the following analyses were performed independently for the OISST and GFDL-CM4C192 data.

\begin{figure}[t]
    \centering
    \noindent\includegraphics[width=16.45cm]{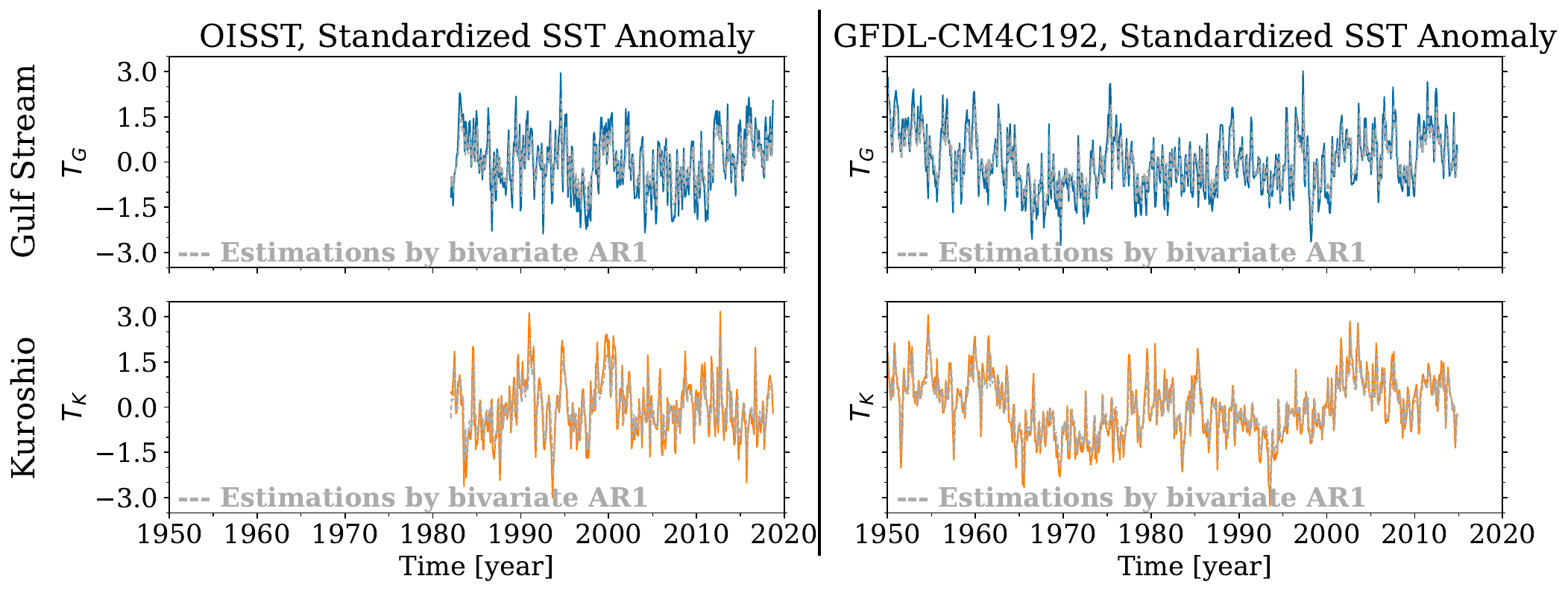}
    \caption{Time series of the standardized SST anomalies for the Gulf Stream (blue) and the Kuroshio Current (orange) from the OISST and GFDL-CM4C192 data. The dashed gray lines represent the regression results for the corresponding time series with the first-order autoregressive (AR1) model.}
    \label{fig:time-series-regression}
\end{figure}

Figure \ref{fig:time-series-regression} shows the OISST data, GFDL-CM4C192 data, and their regression results, where a first-order bivariate autoregressive (AR1) model was employed \citep{Brockwell+Davis91Book,Mudelsee14Book}. In estimating the AR1 model, the SST at the current time step was regressed onto the SST at the previous time step, and the regression coefficients were obtained. This regression method is known as the ordinary least squares (OLS) and is standard in time series analysis \citep{Hamilton94Book,Mudelsee14Book}. The AR1 model reproduces the SST variations well (Fig. \ref{fig:time-series-regression}), suggesting the validity of the AR1 model. The estimated AR1 model can be uniquely converted to the continuous-time dynamical system [Eqs. (\ref{eq:dynamical-sys-TG}) and (\ref{eq:dynamical-sys-TK})]. Specifically, under the assumption of statistical stationarity, we require that the means and variances of the variations in $T_G$ and $T_K$ are the same between the AR1 model and the dynamical system  (see Subsection \ref{subsec:linear-SDE-and-AR1} in Appendix B for details). In other words, this conversion requires that the statistical properties of the AR1 model match those of the time-integrated values over each discrete time step (i.e., one month) using the dynamical system.

We used the moving block bootstrap method \citep{Mudelsee14Book} to quantify the uncertainty in the estimated coefficients of the dynamical system. This bootstrap method is a non-parametric method that does not assume normality. This method first splits the original time series into blocks. These blocks are resampled with replacement, and the resampled blocks are connected to generate a new time series of the same length as the original time series, which is called the resampled time series. The coefficients of the dynamical system are estimated for each resampled time series. We can compute confidence intervals of the coefficients for the dynamical system by repeating this procedure and creating histograms of the coefficients. The block length was quantitatively determined from the auto-correlation coefficients \citep{Sherman+98JAppStats, Mudelsee14Book}: the block lengths were $18$ months for the OISST data and $30$ months for the GFDL-CM4C192 data. The shorter block length for the OISST data may be attributed to its shorter time series length. The number of resamples was set to $2000$ following \citet{Mudelsee14Book}.

Table \ref{table:estimated-coeffs} shows the estimation results of the coefficients of the dynamical system, where the brackets show the 95\% confidence intervals by the bootstrap method. The relaxation coefficients $r_G$ and $r_K$ are positive, and their inverses give the relaxation time scales. The time scale is about $3$ months for the Gulf Stream and the Kuroshio Current. The estimates of the interaction coefficients $c_{G \leftarrow K}$ and $c_{K \leftarrow G}$ are positive. The confidence intervals suggest that $c_{G \leftarrow K}$ based on the OISST data may be negative. The impact of this negative value is discussed in Appendix C. The variances of the noise $D_G$ and $D_K$ are of similar magnitude. We confirmed that the uncertainty quantification is not strongly dependent on the choice of method, as similar results were obtained using a parametric approach based on normality (not shown).

\begin{table}[t]
    \caption{Estimated coefficients of the dynamical system [Eqs. (\ref{eq:dynamical-sys-TG}) and (\ref{eq:dynamical-sys-TK})] from the OISST and GFDL-CM4C192 data. The values represent the estimates obtained from the regression analysis, and the brackets show the 95\% confidence intervals obtained from the moving block bootstrap method.}
    \label{table:estimated-coeffs}
    \begin{center}
        \begin{tabular}{lll}
            \hline\hline
            & OISST & GFDL-CM4C192 \\
            \hline
            $r_G$                 & 0.343  [0.327, 0.555]   & 0.299  [0.289, 0.429]  \\
            $r_K$                 & 0.325  [0.301, 0.538]   & 0.206  [0.196, 0.321]  \\
            $c_{G \leftarrow K}$ & 0.0280 [-0.0634, 0.117] & 0.105  [0.0433, 0.171] \\
            $c_{K \leftarrow G}$ & 0.110  [0.00387, 0.240] & 0.0756 [0.0249, 0.144] \\
            $D_G$                 & 0.343  [0.324, 0.545]   & 0.253  [0.252, 0.382]  \\
            $D_K$                 & 0.308  [0.278, 0.519]   & 0.176  [0.173, 0.283]  \\
            \hline
            \end{tabular}
    \end{center}
\end{table}


In the absence of noise, the solution of the dynamical system decays to zero without oscillation. According to linear dynamical system theory \citep[e.g.,][]{Strogatz18Book}, solutions of a deterministic linear system are categorized using the trace and determinant of its coefficient matrix. In our case, the coefficient matrices from all $2000$ resampled time series exhibit decaying solutions, indicating that the dynamical system is stable. Moreover, most of these estimated coefficients are categorized into the decaying solution without oscillation when $c_{G \leftarrow K} > 0$ and $c_{K \leftarrow G} > 0$: all estimates from the GFDL-CM4C192 data belong to this category, whereas $1624$ out of $2000$ estimates from the OISST data belong to this category. For the OISST, the remaining $376$ estimates show decaying oscillators due to the negative $c_{G \leftarrow K}$ (Table \ref{table:estimated-coeffs}), the case of which is discussed in Appendix C. The influences of noise are analyzed using information thermodynamics in Section \ref{sec:analysis}.

\subsection{Validation of the assumptions for the dynamical system} \label{subsec:valid-dynamical-system}

We evaluate the validity of the estimated dynamical system. The following assumptions have been made regarding the dynamical system: (i) statistical stationarity and (ii) zero auto- and cross-correlations for the noises $\xi_G$ and $\xi_K$. We verify these two assumptions.

First, regarding statistical stationarity, we used the Dickey-Fuller test \citep{Hamilton94Book}, which tests the null hypothesis of non-stationarity against the alternative hypothesis of a stationary AR1 process. For the OISST and GFDL-CM4C192 data, the $p$-value was at most around $10^{-13}$. This result supports the validity of regarding each time series as stationary.

Second, we confirm the validity of no correlations, namely Eqs. (\ref{eq:auto-corr-xi_G})--(\ref{eq:cross-corr-xi_G-xi_K}). The first two equations indicate that the noises do not depend on the past, resulting in zero auto-correlations. Equation (\ref{eq:cross-corr-xi_G-xi_K}) indicates that the noises $\xi_G$ and $\xi_K$ are independent of each other, resulting in zero cross-correlations. In regression analysis, the residuals between the estimates and the target variables are considered as noise. Thus, we can verify the assumptions by examining the auto- and cross-correlation coefficients of these residuals.

\begin{figure}[t]
    \centering
    \noindent\includegraphics[width=16.45cm]{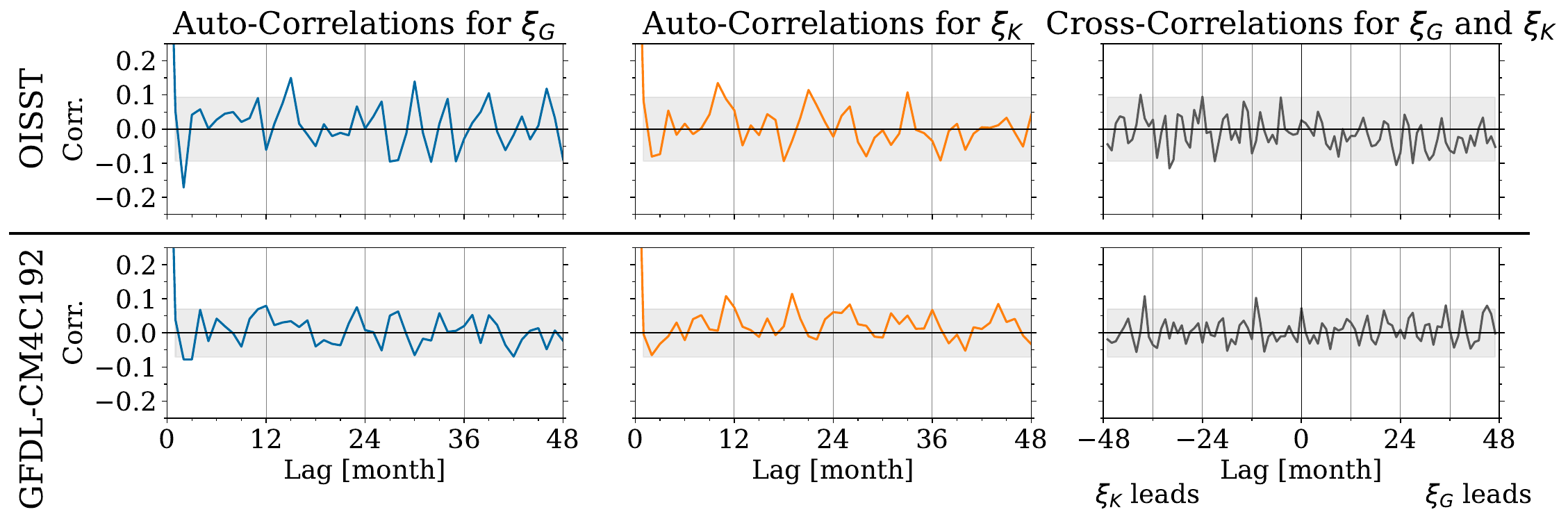}
    \caption{Auto- and cross-correlation coefficients of the residuals between the AR1-model estimates and the target time series for the OISST and GFDL-CM4C192 data. The gray shading indicates the 95\% confidence intervals.}
    \label{fig:correlation-analysis}
\end{figure}

Figure \ref{fig:correlation-analysis} shows the results of the correlation analysis for these residuals. The gray areas represent the 95\% confidence intervals \citep{Brockwell+Davis91Book}. For most lags, the auto-correlation coefficients fall within the confidence intervals, suggesting that the auto-correlations can be considered zero. These results support the assumption that the noises do not strongly depend on the past. The cross-correlation coefficients also take values close to zero and fall within the confidence intervals for most time lags. This result supports the assumption that the noises of the Gulf Stream and the Kuroshio Current are statistically independent. \citet{Kohyama+21Science} showed that the BCS is not a passive phenomenon driven by the atmospheric baroclinic jet. Our results suggest that the SSTs of both currents are not driven by common noise, which is consistent with the result of \citet{Kohyama+21Science}. These noises are attributed to atmospheric and oceanic fluctuations on time scales of a few months or less, likely encompassing the spatially localized effects of the atmospheric baroclinic jet and oceanic mesoscale eddies in the Gulf Stream and Kuroshio Current regions.

We finally clarify the relationship between noise and temperature. In non-equilibrium physics, the variance of stochastic noise is generally proportional to the temperature of an isothermal environment; this noise arises from collisions of small molecules with the larger target particle \citep[e.g.,][]{Sekimoto10Book}. In our case, since the noises $\xi_G$ and $\xi_K$ represent atmospheric and oceanic fluctuations, their variances $D_G$ and $D_K$ are interpreted as the strengths of these small-scale disturbances. Thus, isothermal environments correspond to the disturbance fields, and ``temperature'' in non-equilibrium physics corresponds to $D_G$ and $D_K$. These effective temperatures should be distinguished from the SST anomalies $T_G$ and $T_K$ at longer time scales.

\subsection{Reproduction of the BCS by the dynamical system} \label{subsec:reproduction-bcs}

Finally, we confirm that the estimated dynamical system reproduces the BCS between $T_G$ and $T_K$. To show the BCS, we examine the lag correlation between $T_G$ and $T_K$ following \citet{Kohyama+21Science}, who directly obtained the lag correlation coefficients from the time series. In the present study, we first estimated the coefficients of the dynamical system from the same time series. Then, using these coefficients, we calculated the analytical solutions of the lag correlation coefficients using the theoretical formulas (Subsection \ref{subsec:theory-lag-corr} in Appendix B).

Figure \ref{fig:lag-correlation} shows the resultant lag correlations. Here, the 95\% confidence interval for each lag correlation coefficient was estimated using the $2000$ sets of resampled time series obtained by the moving block bootstrap method. The solid lines indicate that the correlation coefficient takes significant values larger than zero, which supports the existence of the synchronicity. The longer interval of significant correlations for the GFDL-CM4C192 data is likely due to its longer data range compared to that of the OISST data. The Gulf Stream slightly leads the Kuroshio Current by 1 and 0.2 months for the OISST and GFDL-CM4C192 data, respectively. The significance of these small leads is difficult to assess given the short data periods. The obtained lag correlation coefficients are very close to the results of \citet{Kohyama+21Science} (see their Fig. 1D). Therefore, we can conclude that the dynamical system serves as an appropriate model for describing BCS.

\begin{figure}[t]
    \centering
    \noindent\includegraphics[width=11.4cm]{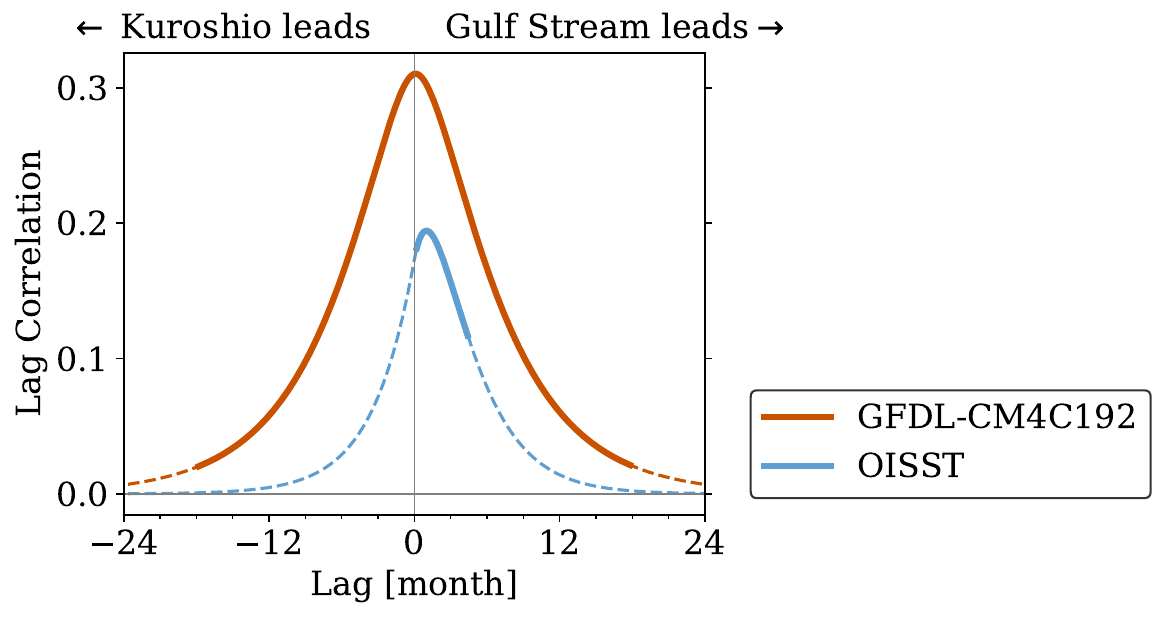}
    \caption{Lag correlation coefficients between the Gulf Stream and Kuroshio Current SSTs for the OISST and GFDL-CM4C192 data. The solid lines indicate significant correlations at the 95\% confidence level, while the dashed lines represent insignificant correlations.}
    \label{fig:lag-correlation}
\end{figure}

\section{Analysis of the dynamical system based on information thermodynamics} \label{sec:analysis}

\subsection{Application of the second law of information thermodynamics} \label{subsec:application-second-law}

We apply the second law of information thermodynamics \citep{Horowitz+Esposito14PhysRevX,Loos+Klapp20NewJPhys} to the Gulf Stream $T_G$ and the Kuroshio Current $T_K$ to clarify their asymmetric roles in the BCS. We assume a statistical steady state in which the probability distribution of $T_G$ and $T_K$ does not vary over time. The time-series analysis supports the validity of the steady-state assumption (Section \ref{sec:dynamical-system}\ref{subsec:valid-dynamical-system}).

The second law of information thermodynamics gives the following inequalities and equality regarding the information flows and entropy production rates for the Gulf Stream and the Kuroshio Current \citep{Horowitz+Esposito14PhysRevX,Loos+Klapp20NewJPhys}:
\begin{align}
\dot{\sigma}_G &\ge \dot{I}_{G \leftarrow K}, \label{eq:second-law-TG}\\
\dot{\sigma}_K &\ge \dot{I}_{K \leftarrow G}, \label{eq:second-law-TK}\\
0 &= \dot{I}_{G \leftarrow K} + \dot{I}_{K \leftarrow G}, \label{eq:conservation-info-flow-TG-TK}
\end{align}
where $\dot{\sigma}_G$ and $\dot{\sigma}_K$ represent the production rates of total entropy for the Gulf Stream $T_G$ and the Kuroshio Current $T_K$, respectively, and $\dot{I}_{G \leftarrow K}$ and $\dot{I}_{K \leftarrow G}$ represent the information flow to the Gulf Stream and to the Kuroshio Current, respectively. Equations (\ref{eq:second-law-TG}), (\ref{eq:second-law-TK}), and (\ref{eq:conservation-info-flow-TG-TK}) correspond to Eqs. (\ref{eq:second-law-info-autonomousX}), (\ref{eq:second-law-info-autonomousY}), and (\ref{eq:conservation-I-X-Y}), respectively. The sum of the two information flows is zero in Eq. (\ref{eq:conservation-info-flow-TG-TK}) because the time derivative of mutual information is zero due to the steady-state assumption (Appendix A). Moreover, under this assumption, the production rates of total entropy correspond to $\Delta Q$ in Eq. (\ref{eq:total-entropy-X}) because the change of Shannon entropies is zero. The interpretation of $\Delta Q$ for the dynamical system is discussed in Subsection \ref{subsec:interpretation-heat} in Appendix A. Adding Eqs. (\ref{eq:second-law-TG}) and (\ref{eq:second-law-TK}) yields $\dot{\sigma}_G + \dot{\sigma}_K \ge 0$, i.e., the total entropy production for the entire system is non-negative, indicating the consistency with the second law of thermodynamics. Considering each subsystem, $\dot{\sigma}_G$ and  $\dot{\sigma}_K$ are lower bounded by $\dot{I}_{G \leftarrow K}$ and $\dot{I}_{K \leftarrow G}$, respectively, one of which can be negative. By using information thermodynamics, we investigate the asymmetry between the dynamics of $T_G$ and $T_K$ from the perspective of entropy production and information transfer.

For the dynamical system [Eqs. (\ref{eq:dynamical-sys-TG}) and (\ref{eq:dynamical-sys-TK})], the analytical solutions for the entropy production rates and information flows are obtained \citep{Loos+Klapp20NewJPhys}:
\begin{align}
\dot{\sigma}_G &= \frac{c_{G \leftarrow K}}{D_G} \left\langle T_K \frac{{\rm d}T_G}{{\rm d}t} \right\rangle,\label{eq:sigma_G}\\
\dot{\sigma}_K &= \frac{c_{K \leftarrow G}}{D_K} \left\langle T_G \frac{{\rm d}T_K}{{\rm d} t}\right\rangle,\label{eq:sigma_K}\\
\dot{I}_{G \leftarrow K} &= \frac{\Sigma_{GK}}{\lvert \bm{\mathsf{\Sigma}} \rvert} \left\langle T_K  \frac{{\rm d}T_G}{{\rm d}t} \right\rangle,\label{eq:I_G}\\
\dot{I}_{K \leftarrow G} &= \frac{\Sigma_{GK}}{\lvert \bm{\mathsf{\Sigma}} \rvert} \left\langle T_G \frac{{\rm d}T_K}{{\rm d} t} \right\rangle, \label{eq:I_K}
\end{align}
where we introduce the covariance matrix $\bm{\mathsf{\Sigma}}$ between $T_G$ and $T_K$. The off-diagonal component of $\bm{\mathsf{\Sigma}}$ is represented by $\Sigma_{GK}$, and the determinant of $\bm{\mathsf{\Sigma}}$ is represented by $\lvert \bm{\mathsf{\Sigma}} \rvert$. Equations (\ref{eq:sigma_G})--(\ref{eq:I_K}) indicate that the entropy production rates and information flows are linked to the correlations between the SSTs and their rates of change. Thus, the temporal characteristics of the dynamical system can be examined using information thermodynamics. Details on the derivation of Eqs. (\ref{eq:sigma_G})--(\ref{eq:I_K}) are given in Subsection \ref{subsec:theory-entropy-info-flow} in Appendix B.

In the steady state, the entropy production rate and information flow of the Kuroshio Current are related to those of the Gulf Stream because of the following relationship (Subsection \ref{subsec:theory-entropy-info-flow} in Appendix B):
\begin{equation}
    \left\langle T_K \frac{{\rm d}T_G}{{\rm d}t} \right\rangle + \left\langle T_G \frac{{\rm d}T_K}{{\rm d}t} \right\rangle = 0. \label{eq:correlation-TG-TK-in-steady-state}
\end{equation}
Applying Eq. (\ref{eq:correlation-TG-TK-in-steady-state}) into Eqs. (\ref{eq:sigma_G})--(\ref{eq:I_K}), we obtain the following relationships:
\begin{align}
     \dot{\sigma}_K &= -\frac{c_{K \leftarrow G} D_G}{c_{G \leftarrow K} D_K} \dot{\sigma}_G, \label{eq:relation-sigmaG-sigmaK} \\
     \dot{I}_{K \leftarrow G} &= -\dot{I}_{G \leftarrow K}. \label{eq:relation-IG-IK}
\end{align}
Equation (\ref{eq:relation-IG-IK}) can also be derived from the conservation of the information flows [Eq. (\ref{eq:conservation-info-flow-TG-TK})]. We analyze the dynamical system using the theoretical formulas (\ref{eq:sigma_G})--(\ref{eq:relation-IG-IK}).

\subsection{Analysis of regime diagrams for the BCS} \label{subsec:regime-diagrams}

\begin{figure}[t]
    \centering
    \noindent\includegraphics[width=16.45cm]{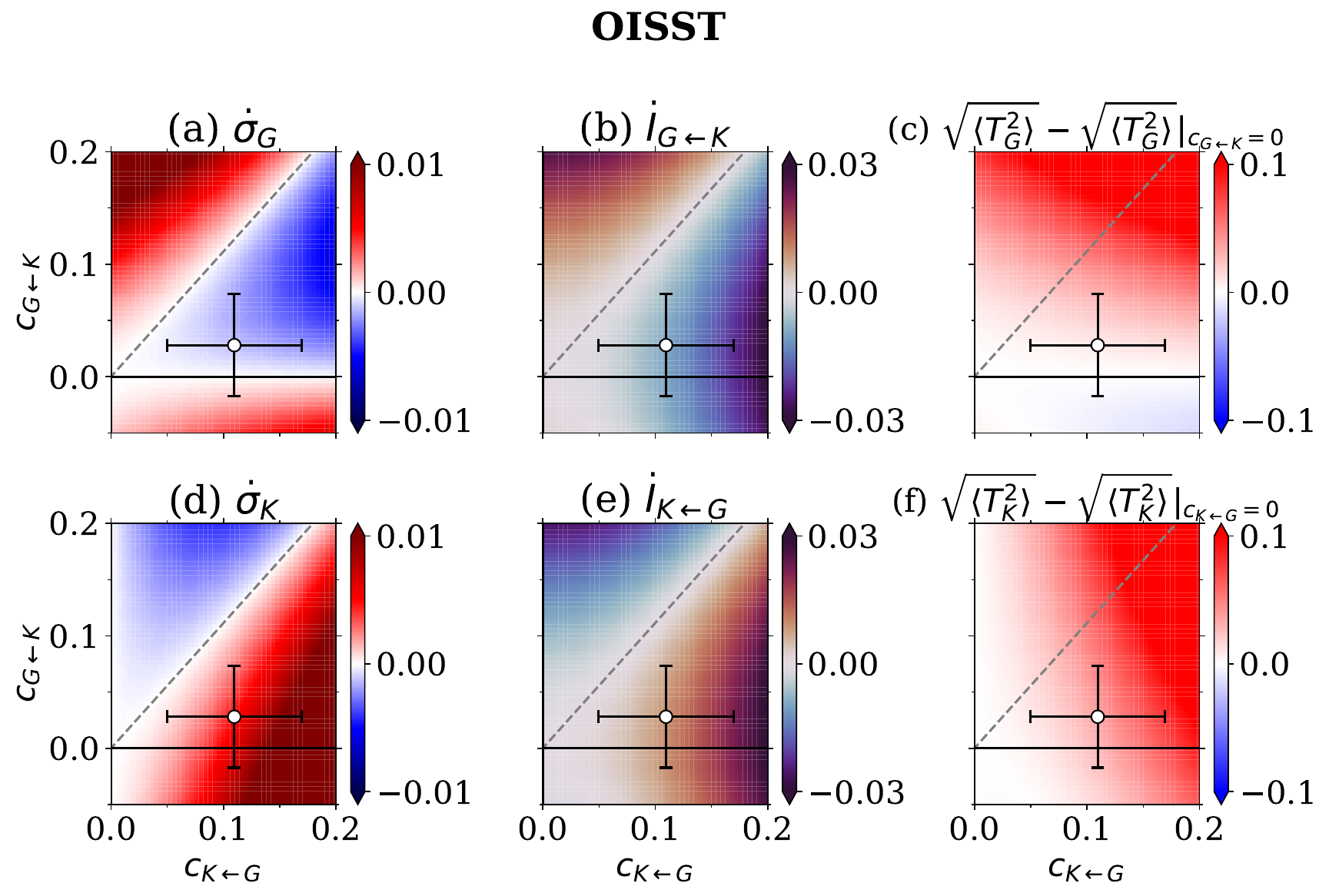}
    \caption{Dependence of information thermodynamic quantities on the interaction coefficients $c_{K \leftarrow G}$ and $c_{G \leftarrow K}$ for the OISST data: (a) $\dot{\sigma}_G$, (b) $\dot{I}_{G \leftarrow K}$, (c) $\sqrt{\langle T_G^2 \rangle} - \sqrt{\langle T_G^2 \rangle}|_{c_{G \leftarrow K} = 0}$, (d) $\dot{\sigma}_K$, (e) $\dot{I}_{K \leftarrow G}$, and (f) $\sqrt{\langle T_K^2 \rangle} - \sqrt{\langle T_K^2 \rangle|}_{c_{K \leftarrow G} = 0}$. Sub-figures (c) and (f) show the differences in the standard deviations of the SSTs from those without the interactions, namely, when $c_{G \leftarrow K} = 0$ and $c_{K \leftarrow G} = 0$, respectively. The white dots represent the estimated values of $(c_{K \leftarrow G}, c_{G \leftarrow K})$, whose values are listed in Table \ref{table:estimated-coeffs}. The error bars indicate the standard deviations obtained from the bootstrap method. The gray dashed diagonal lines show the symmetric condition $c_{G \leftarrow K} D_K = c_{K \leftarrow G} D_G$.}
    \label{fig:regime-diagram-OISST}
\end{figure}

\begin{figure}[t]
    \centering
    \noindent\includegraphics[width=16.45cm]{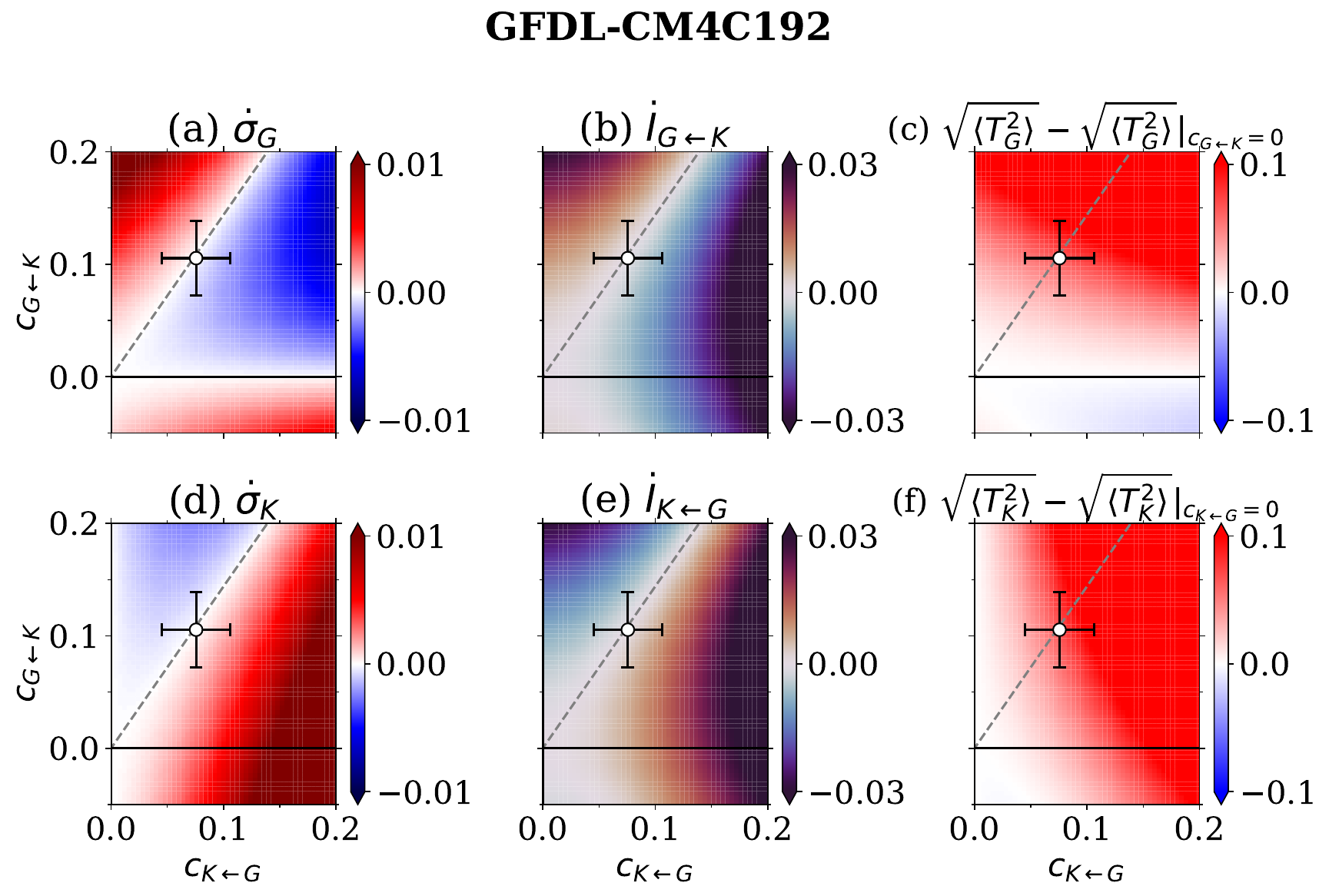}
    \caption{Same as Fig. \ref{fig:regime-diagram-OISST} but for the GFDL-CM4C192 data.}
    \label{fig:regime-diagram-GFDL}
\end{figure}

We show that the dynamical system can be regarded as a Maxwell's demon system by examining the dependence of information thermodynamics quantities on the interaction coefficients $c_{K \leftarrow G}$ and $c_{G \leftarrow K}$ shown in Figs. \ref{fig:regime-diagram-OISST} and \ref{fig:regime-diagram-GFDL}. The parameters $c_{K \leftarrow G}$ and $c_{G \leftarrow K}$ are the most important because they characterize the interactions between the Gulf Stream and the Kuroshio Current. In Figs. \ref{fig:regime-diagram-OISST} and \ref{fig:regime-diagram-GFDL}, the white dots represent the estimated values of $(c_{K \leftarrow G},c_{G \leftarrow K})$ listed in Table \ref{table:estimated-coeffs}. The error bars indicate the standard deviations obtained by the bootstrap method.

The error bars for the estimated values suggest that the true value is likely to lie in the region where $c_{K \leftarrow G} > 0$ and $c_{G \leftarrow K} > 0$ (Figs. \ref{fig:regime-diagram-OISST} and \ref{fig:regime-diagram-GFDL}), though the results based on the OISST data suggest that the true value may possibly lie in the region where $c_{G \leftarrow K} < 0$. Since the period of the OISST data is shorter, the uncertainty in the estimation is larger compared to the results of the GFDL-CM4C192 data. Appendix C briefly discusses the case where $c_{G \leftarrow K} < 0$. Hereafter, we focus on the region where $c_{G \leftarrow K} > 0$.

The Gulf Stream and the Kuroshio Current are symmetric on the gray dashed diagonal lines in Figs. \ref{fig:regime-diagram-OISST} and \ref{fig:regime-diagram-GFDL}, where $c_{G \leftarrow K} D_K = c_{K \leftarrow G} D_G$ (Appendix D). On these lines, the interchange of $T_G$ and $T_K$ does not affect the entropy production rates and information flows because these quantities are zero \citep{Loos+Klapp20NewJPhys}. Below these lines, $\dot{\sigma}_G$ and $\dot{I}_{G \leftarrow K}$ are negative for the Gulf Stream, whereas $\dot{\sigma}_K$ and $\dot{I}_{K \leftarrow G}$ are positive for the Kuroshio Current. Above these lines, all the signs of these quantities are reversed. Thus, the roles of the Gulf Stream and the Kuroshio Current are reversed across the diagonal lines in Figs. \ref{fig:regime-diagram-OISST} and \ref{fig:regime-diagram-GFDL}.

The Gulf Stream and the Kuroshio Current are likely to be asymmetric due to various factors, such as the land-sea distribution. This consideration suggests that the true value is not strictly on the gray dashed diagonal lines in Figs. \ref{fig:regime-diagram-OISST} and \ref{fig:regime-diagram-GFDL}, although the estimated values from the GFDL-CM4C192 data are quite close to these diagonal lines (Fig. \ref{fig:regime-diagram-GFDL}). This result is probably due to the fact that small-scale disturbances regarded as noise are not sufficiently reproduced in the GFDL-CM4C192 model. Indeed, the noise variances $D_G$ and $D_K$ are smaller than those estimated from the OISST data (Table \ref{table:estimated-coeffs}), and these parameters affect the symmetric condition $c_{G \leftarrow K} D_K = c_{K \leftarrow G} D_G$. Hereafter, we consider that the true value exists on the lower side of the diagonal lines, where both estimated values from the OISST and GFDL-CM4C192 data belong.

Figures \ref{fig:regime-diagram-OISST} and \ref{fig:regime-diagram-GFDL} indicate that the dynamical system is a Maxwell's demon system, where the Gulf Stream is a ``Particle'' and the Kuroshio Current is a ``Demon.'' The entropy production rate and information flow of the Gulf Stream, $\dot{\sigma}_G$ and $\dot{I}_{G \leftarrow K}$, are negative  (Figs. \ref{fig:regime-diagram-OISST}a, \ref{fig:regime-diagram-OISST}b, \ref{fig:regime-diagram-GFDL}a, and \ref{fig:regime-diagram-GFDL}b), whereas those of the Kuroshio Current, $\dot{\sigma}_K$ and $\dot{I}_{K \leftarrow G}$, are positive (Figs. \ref{fig:regime-diagram-OISST}d, \ref{fig:regime-diagram-OISST}e, \ref{fig:regime-diagram-GFDL}d, and \ref{fig:regime-diagram-GFDL}e). According to the discussions on Maxwell's demon systems (Sections \ref{sec:maxwell-demon}\ref{subsec:resolution-maxwell-demon} and \ref{sec:maxwell-demon}\ref{subsec:autonomous-maxwell-demon}), the Kuroshio Current as the ``Demon'' tends to establish the correlation with the Gulf Stream SST by measuring it, which is reflected in the positive information flow ($\dot{I}_{K \leftarrow G} > 0$). Furthermore, the Kuroshio Current utilizes its own SST as a memory and exhibits a positive entropy production rate ($\dot{\sigma}_K > 0$), which is attributed to the memory update process associated with the measurement. On the other hand, the Gulf Stream as the ``Particle'' tends to decrease the correlation due to the feedback control by the Kuroshio Current, which is reflected in the negative information flow ($\dot{I}_{G \leftarrow K} < 0$). In addition, the Gulf Stream SST varies by rectifying noise and shows a negative entropy production rate ($\dot{\sigma}_G < 0$). The term ``rectify'' means to extract coherent directional output from random noise fluctuations.

We further discuss the rectification of noise. Figures \ref{fig:regime-diagram-OISST}c, \ref{fig:regime-diagram-OISST}f, \ref{fig:regime-diagram-GFDL}c, and \ref{fig:regime-diagram-GFDL}f show that the SST magnitudes $\sqrt{\langle T_G^2 \rangle}$ and $\sqrt{\langle T_K^2 \rangle}$ are increased compared to the case without interactions. As discussed in Section \ref{sec:dynamical-system}\ref{subsec:regression-analysis}, without noise, the dynamical system would decay to zero without oscillation. This result suggests that the driving source is noise. Moreover, in the absence of interactions, namely $c_{G \leftarrow K} = c_{K \leftarrow G} = 0$, the SSTs of the Gulf Stream and the Kuroshio Current would evolve independently, and BCS would not occur. Compared to the SST magnitudes in this case, they are increased due to the interactions. Within the framework of the dynamical system, the presence of both interactions and noises is essential for BCS.

We first investigate the reason for the increase in $T_G$. The entropy production rate $\dot{\sigma}_G$ is given as follows, which is negative (Figs. \ref{fig:regime-diagram-OISST}a and \ref{fig:regime-diagram-GFDL}a):
\begin{equation}
    0 > \dot{\sigma}_G = \frac{c_{G \leftarrow K}}{D_G} \left\langle T_K \frac{{\rm d}T_G}{{\rm d} t} \right\rangle = \frac{(c_{G \leftarrow K})^2}{D_G}\langle T_K T_K\rangle - \frac{r_G c_{G \leftarrow K}}{D_G} \langle T_K T_G\rangle, \label{eq:sigma_G_decomposition}
\end{equation}
where all coefficients are positive. The rightmost side is obtained by substituting Eq. (\ref{eq:dynamical-sys-TG}) into (\ref{eq:sigma_G}). The relative magnitudes of the two terms on the rightmost side characterize the behavior of the Gulf Stream. When the first term dominates, indicating a stronger influence from the Kuroshio Current, the Gulf Stream shows increased fluctuations and a larger entropy production rate $\dot{\sigma}_G$. Conversely, when the second term dominates, the relaxation is more effective, reducing the stochasticity of the Gulf Stream (i.e., decreasing $\dot{\sigma}_G$). Clearly, the negative $\dot{\sigma}_G$ indicates that the second term is dominant.

\begin{figure}[t]
    \centering
    \noindent\includegraphics[width=16.45cm]{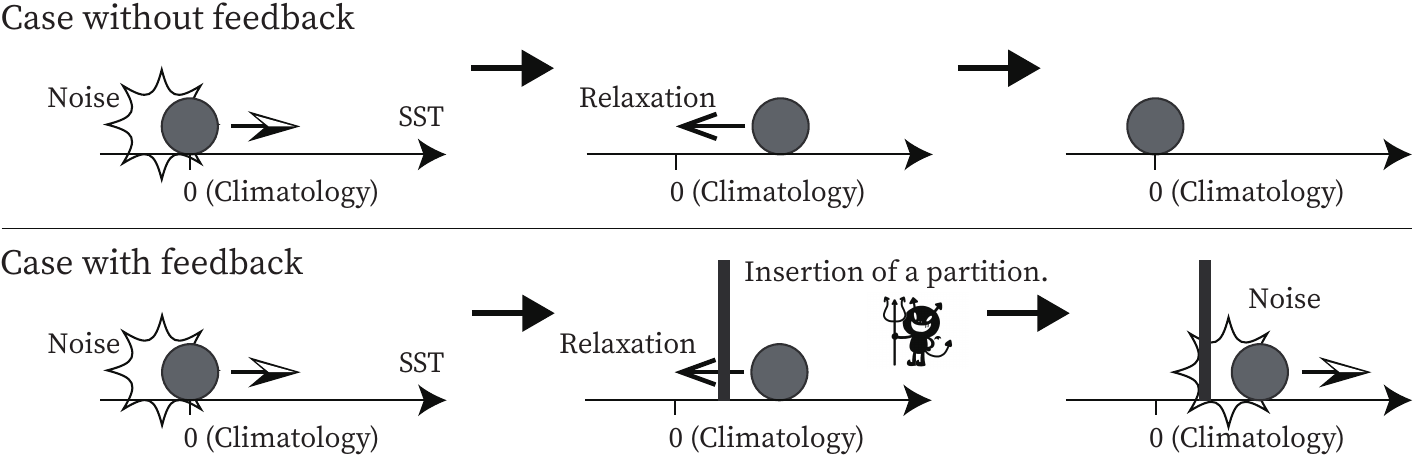}
    \caption{Schematic of the increase in the Gulf Stream SST due to noise and feedback. Without the action from the Kuroshio Current, namely no feedback (top), the SST increases due to noise and quickly returns to zero through relaxation. With feedback from the Kuroshio Current (bottom), relaxation is interfered with, allowing the possibility of further increase by noise before returning to zero. The Kuroshio Current measures the Gulf Stream SST and inserts a ``partition'' at an appropriate location.}
    \label{fig:feedback-control-sst}
\end{figure}

The negative $\dot{\sigma}_G$ suggests that the action from the Kuroshio Current on the Gulf Stream, $c_{G \leftarrow K} T_K$, does not directly contribute to the increase in $T_G$. Indeed, this action $c_{G \leftarrow K} T_K$ is negatively correlated with the tendency of the Gulf Stream SST, $c_{G \leftarrow K}\langle T_K \, {\rm d}T_G/{\rm d}t\rangle = D_G \dot{\sigma}_G < 0$. This result is due to the relaxation. Without relaxation, the governing equation (\ref{eq:dynamical-sys-TG}) would be ${\rm d}T_G/{\rm d}t = c_{G \leftarrow K} T_K$, and only the first term in Eq. (\ref{eq:sigma_G_decomposition}) would remain, leading to a positive $\dot{\sigma}_G$. The mechanism for the increase in $T_G$ is then described as follows. For example, when $T_G$ is positive, it tends to decrease to zero due to relaxation, but this decrease is mitigated by $c_{G \leftarrow K} T_K$. In other words, the action from the Kuroshio Current suppresses the relaxation of the Gulf Stream. Consequently, the remaining term $\xi_G$ in Eq. (\ref{eq:dynamical-sys-TG}), namely the noise, becomes more effective and increase $T_G$. This suppression of relaxation to zero corresponds to the feedback control that inserts a partition (Fig. \ref{fig:feedback-control-sst}).

We next investigate the reason for the increase in $T_K$. The expression for the entropy production rate $\dot{\sigma}_K$ is given as follows, which is positive (Figs. \ref{fig:regime-diagram-OISST}d and \ref{fig:regime-diagram-GFDL}d):
\begin{equation}
    0 < \dot{\sigma}_K = \frac{c_{K \leftarrow G}}{D_K} \left\langle T_G \frac{{\rm d}T_K}{{\rm d}t} \right\rangle = \frac{\left(c_{K \leftarrow G}\right)^2}{D_K} \langle  T_G T_G \rangle - \frac{r_K c_{K \leftarrow G}}{D_K} \langle T_G T_K \rangle, \label{eq:sigma_K_decomposition}
\end{equation}
where all coefficients are positive. The rightmost side is obtained by substituting Eq. (\ref{eq:dynamical-sys-TK}) into (\ref{eq:sigma_K}). Equation (\ref{eq:sigma_K_decomposition}) is interpreted in the same manner as Eq. (\ref{eq:sigma_G_decomposition}). When the first term dominates, the Kuroshio Current shows strong fluctuations and a larger entropy production rate $\dot{\sigma}_K$. Conversely, when the second term dominates, the stochasticity of the Kuroshio Current is reduced and $\dot{\sigma}_K$ decreases. The result of $\dot{\sigma}_K > 0$ indicates the dominance of the first term.

The positive $\dot{\sigma}_K$ suggests that the action from the Gulf Stream on the Kuroshio Current, $c_{K \leftarrow G} T_G$, directly contributes to the increase in $T_K$. Similar to the Gulf Stream, there is an effect of suppressing relaxation; however, its contribution is relatively small and $\dot{\sigma}_K$ is positive. The amplification of $T_K$ occurs by following the changes in $T_G$ due to the action $c_{K \leftarrow G} T_G$ from the Gulf Stream. According to the framework of Maxwell's demon systems, the Kuroshio Current (i.e., the ``Demon'') uses its SST as memory and the SST fluctuations are increased by updating the memory value $T_K$ based on the measurement outcome of the Gulf Stream SST.

These analyses of $T_G$ and $T_K$ suggest that information-to-energy conversion \citep{Toyabe+2010NatPhys} occurs in a subsystem (i.e., the Gulf Stream), even though the entire system is dissipative, as in other Maxwell demon systems. Generally, stochastic noise fluctuates a system back and forth, so the net energy or work is difficult to extract. However, using the information from measurements, fluctuations can be converted into directional movements in a subsystem \citep{Toyabe+2010NatPhys}, corresponding to the negative $\dot{\sigma}_G$ or increased $T_G$ amplitude in our system. Such conversion cannot occur for the entire dissipative system (i.e., no net positive conversion), as reflected in the positive value of $\dot{\sigma}_G + \dot{\sigma}_K$.

Synchronicity is essential for this conversion. For instance, in Fig. \ref{fig:feedback-control-sst}, the partition needs to be inserted just behind the ``Particle.'' This insertion is based on the memory value of the ``Demon,'' as in the Szilard engine (Fig. \ref{fig:szilard-demon-particle}), where the memory value becomes equal to the particle position after the measurement. Although it is difficult to directly compare our dynamical system to the Szilard engine because our system is autonomous (Section \ref{sec:maxwell-demon}\ref{subsec:autonomous-maxwell-demon}), the memory value $X$ and particle position $Y$ correspond to $T_K$ and $T_G$, respectively. Both subsystems $T_K$ and $T_G$ are influenced by the small-scale disturbance fields (Section \ref{sec:dynamical-system}\ref{subsec:valid-dynamical-system}), corresponding to isothermal environments in the Szilard engine. The ``Demon'' ($T_K$) follows a change in the ``Particle'' ($T_G$), thereby enabling feedback control to convert fluctuations. The $T_K$ amplitude increases due to this following (i.e., its continual updates based on $T_G$), although $T_K$ is dissipative (i.e., $\dot{\sigma}_K > 0$). In the Szilard engine, this dissipation corresponds to the work required to initialize the memory.

\begin{figure}[t]
    \centering
    \noindent\includegraphics[width=11.4cm]{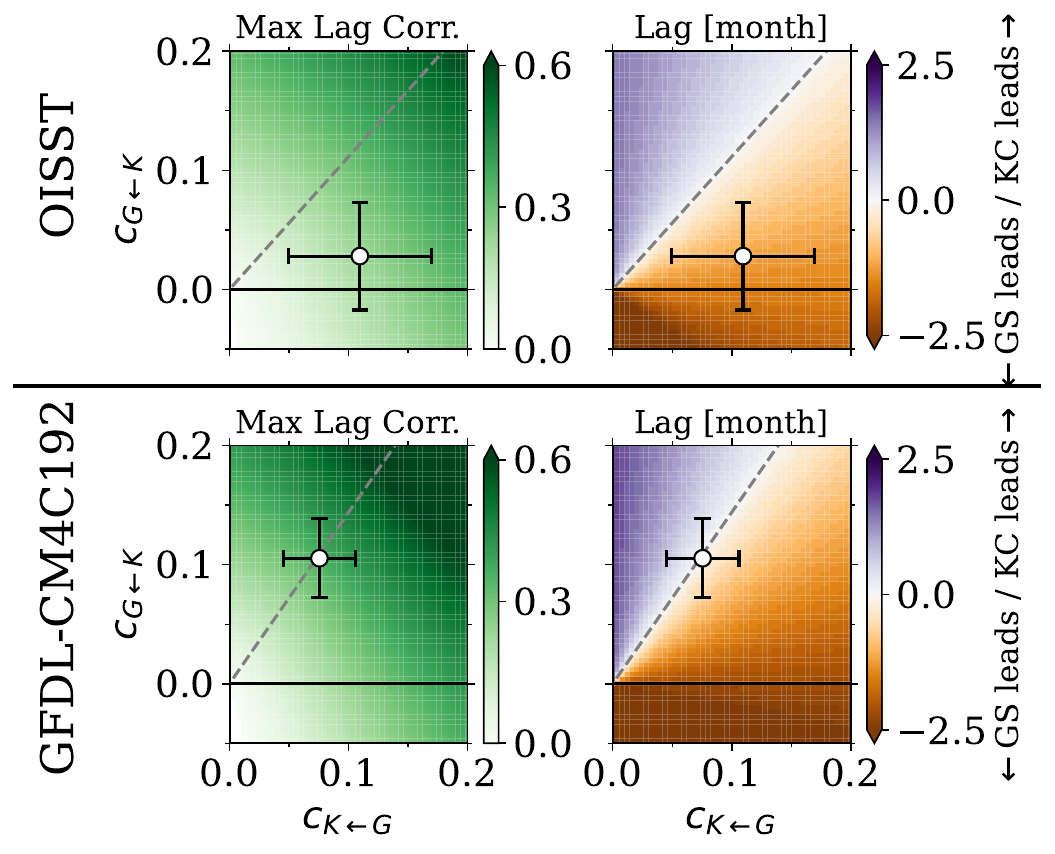}
    \caption{Maximum lag correlation coefficients and the corresponding lags between the Gulf Stream and Kuroshio Current SSTs as functions of the interaction coefficients $c_{K \leftarrow G}$ and $c_{G \leftarrow K}$. The results are shown for the OISST and GFDL-CM4C192 data. The gray dashed diagonal lines show the symmetric condition $c_{G \leftarrow K} D_K = c_{K \leftarrow G} D_G$. The label ``GS leads'' means that the Gulf Stream leads, whereas the label ``KC leads'' means that the Kuroshio Current leads.}
    \label{fig:regime-diagram-corr}
\end{figure}

The dependence of the lag correlation is consistent with the description of the Kuroshio Current following the Gulf Stream. Figure \ref{fig:regime-diagram-corr} shows the dependence of the maximum lag correlation coefficient and the lag at that time over the same parameter space as in Figs. \ref{fig:regime-diagram-OISST} and \ref{fig:regime-diagram-GFDL}. Note that these lag correlation coefficients are theoretically computed as in Fig. \ref{fig:lag-correlation}. As confirmed in Section \ref{sec:dynamical-system}\ref{subsec:reproduction-bcs}, the lag correlation is positive in Fig. \ref{fig:regime-diagram-corr}, suggesting that the BCS occurs. Above the gray dashed diagonal lines in Fig. \ref{fig:regime-diagram-corr}, the SST of the Kuroshio Current leads by $1$--$2$ months, while below the lines, the SST of the Gulf Stream leads. Across the diagonal lines, the roles of the ``Particle'' and the ``Demon'' are reversed because of the reversal of all signs of the entropy production rates and information flows. In the region below the diagonal lines, which we have been considering so far, the Gulf Stream (i.e., the ``Particle'') leads. The Kuroshio Current (i.e., the ``Demon'') measures the Gulf Stream SST and follows the changes in $T_G$.

\subsection{Interpretation of BCS as a Maxwell's demon system: stochastic synchronization} \label{subsec:interpretation-bcs}

\begin{figure}[t]
    \centering
    \noindent\includegraphics[width=14cm]{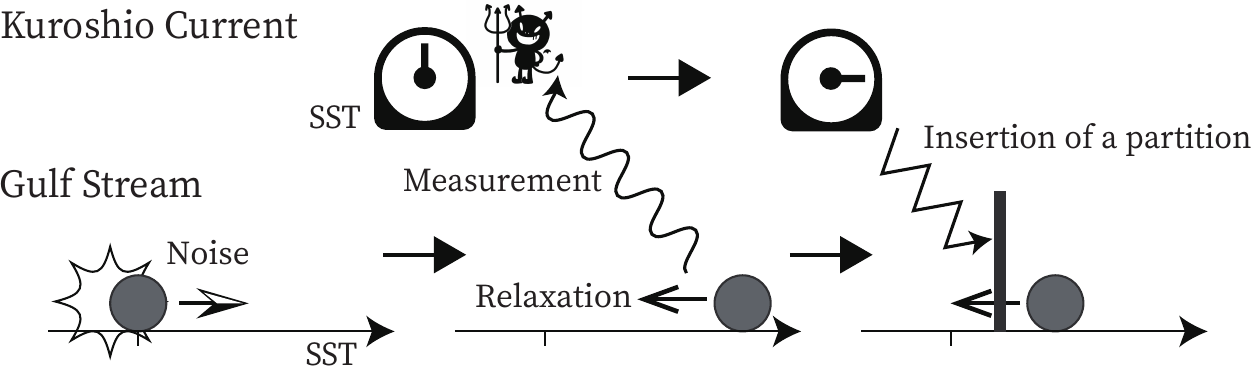}
    \caption{Schematic of the interpretation of the BCS as a Maxwell's demon system (i.e., the stochastic synchronization). The Kuroshio Current plays the role of the ``Demon'' and the Gulf Stream plays the role of the ``Particle.'' See Section \ref{sec:analysis}\ref{subsec:interpretation-bcs} for details.}
    \label{fig:schematic-bcs-maxwell-demon}
\end{figure}

From the above results, we obtain an interpretation of the BCS as a Maxwell's demon system (Fig. \ref{fig:schematic-bcs-maxwell-demon}), which implies the asymmetric roles of the Gulf Stream and the Kuroshio Current. First, the Gulf Stream forces the SST of the Kuroshio Current to be in phase. This effect of making the Kuroshio Current follow is represented by the term $c_{K \leftarrow G} T_G$ in the dynamical system, Eq. (\ref{eq:dynamical-sys-TK}). In the framework of Maxwell's demon, the Gulf Stream is interpreted as being measured by the Kuroshio Current. By contrast, the Kuroshio Current locks the phase of the Gulf Stream SST by interfering with its relaxation toward the climatology. This effect of interference with the relaxation is represented by the term $c_{G \leftarrow K} T_K$ in the dynamical system, Eq. (\ref{eq:dynamical-sys-TG}). In the framework of Maxwell's demon, the Kuroshio Current is interpreted as performing feedback control on the Gulf Stream. On one hand, for the dynamical system without noise, the SSTs would relax to zero without oscillating. On the other hand, without interactions between the Gulf Stream and the Kuroshio Current, their SSTs would fluctuate independently. When both currents are coupled in an appropriate parameter regime, synchronization is realized with atmospheric and oceanic noise as the driving source. The physical origins of this driving source can be attributed to the effects of atmospheric jet streams and oceanic mesoscale eddies on time scales of a few months or less.

In this regard, \cite{Yamagami+2024AOGS} recently examined the asymmetric roles between the Gulf Stream and Kuroshio Current using a coupled atmosphere-ocean general circulation model through ``pacemaker'' experiments. In these experiments, SST variations only in a western boundary current region are strongly relaxed toward (i.e., keep ``pace'' with) a control run, so that the responses of the other climate systems to the western boundary current variability are isolated.  They showed that, when the Gulf Stream acts as a pacemaker, BCS is reproduced. The heat release from the Gulf Stream modulates the atmospheric annular mode, exciting oceanic Rossby waves in the Pacific and influencing the SST in the Kuroshio Current region. This mechanism serves as a plausible physical process that realizes the role of the ``Particle'' in the context of the Maxwell's demon system, i.e., the Gulf Stream forces the SST of the Kuroshio Current to be in phase. Conversely, they also showed that, when the Kuroshio Current acts as a pacemaker, the BCS is not reproduced. This result is also consistent with our interpretation that, unlike the Gulf Stream, the Kuroshio Current does not force the SST of the Gulf Stream by itself. Rather, by responding to the Gulf Stream's forcing, the Kuroshio Current returns a feedback to interfere with the Gulf Stream SST relaxation toward its climatology. The absence of BCS in the Kuroshio pacemaker experiment can be understood that, because its SST varies independently from the state of the Gulf Stream, the Kuroshio Current cannot reproduce an optimal feedback in an appropriate timing. Our future work is to reveal what kind of physical processes regarding the Kuroshio Current realizes the role of the ``Demon'' in the real world.

Our proposed mechanism can be realized in other stochastic dissipative systems with positive feedback. The dynamical system investigated here, Eqs. (\ref{eq:dynamical-sys-TG}) and (\ref{eq:dynamical-sys-TK}), is known as the Langevin system, one of the simplest and most common models with dissipation and stochastic forcing \citep[e.g.,][]{Sekimoto10Book}. Indeed, Langevin systems have been employed as simple models for climate phenomena \citep[e.g.,][]{Dijkstra2013Book}. The key ingredient in our study is the positive feedback; that is, both interaction coefficients $c_{K \leftarrow G}$ and $c_{G \leftarrow K}$ are positive. Due to this positive feedback, atmospheric or oceanic fluctuations are converted into directional variations in a subsystem via synchronization (i.e., information-to-energy conversion; \citet{Toyabe+2010NatPhys}), although such conversion cannot be realized for the entire dissipative system. The wide applicability of Langevin systems suggests that the proposed mechanism of synchronization (Fig. \ref{fig:schematic-bcs-maxwell-demon}), ``stochastic synchronization,'' may describe other climate phenomena. Such applications remain an important topic for future work.

\section{Conclusions} \label{sec:conclusions}

This study has introduced a bivariate linear dynamical system that describes the BCS and analyzed it using a theory of information thermodynamics \citep{Horowitz+Esposito14PhysRevX,Loos+Klapp20NewJPhys}. This system can be interpreted as a Maxwell's demon system, with the Gulf Stream playing the role of the ``Particle'' and the Kuroshio Current playing the role of the ``Demon.'' Information thermodynamics clarifies the asymmetric roles of the Gulf Stream and the Kuroshio Current from the perspective of entropy production and information transfer (Section \ref{sec:analysis}\ref{subsec:interpretation-bcs}). Our results suggest that random noise fluctuations can be autonomously converted into coherent, directional variations (i.e., information-to-energy conversion), through synchronization between the Gulf Stream and the Kuroshio Current. The proposed mechanism, named ``stochastic synchronization," may describe other climate phenomena because our analysis was based on a widely applicable Langevin system.

For future research, we propose applying information thermodynamics to more realistic models that describe the BCS \citep[e.g.,][]{Gallego+Cessi01JClim, Kohyama+21Science}. While the current study demonstrates that the Gulf Stream can be interpreted as the ``Particle'' and the Kuroshio Current as the ``Demon,'' what determines these roles remains unclear, and these roles might be reversed under certain conditions, such as due to temporal changes in interaction coefficients. Moreover, it is unclear how the interactions, such as the interference of the relaxation, are realized in the real climate system. These points will be revealed by applying the theory to realistic models.

Information thermodynamics is applicable to other climate phenomena, such as the El Ni\~{n}o-Southern Oscillation, and may provide new physical insights. However, the direct application may not be straightforward because information thermodynamics requires the estimation of probability, and such estimation is generally difficult for systems with large degrees of freedom. Recently developed data-driven methods, such as dimensionality reduction \citep{Reddy+20IEEE,Hou+22DSE}, may be effective for the probability estimation and expand the applicability of information thermodynamics.

%

%

\clearpage


\acknowledgments

During the preparation of this work, the authors used Claude 3 only for English editing. After using this service, the authors reviewed and edited the content as needed. The authors take full responsibility for the content of the publication. The second author is supported by Japan Society for the Promotion of Science (JSPS) Kakenhi (22H04487, 23H01241, and 23K13169), and the MEXT program for the advanced studies of climate change projection (SENTAN) Grant Number JPMXD0722680395. The authors thank Masaru Inatsu for initiating our connection, leading to this collaboration.

%
%
\datastatement

The data and source code that support the findings of this study are preserved at the Zenodo repository (\url{https://doi.org/10.5281/zenodo.13085327}) and developed openly at the GitHub repository (\url{https://github.com/YukiYasuda2718/bcs_maxwell_demon}).



\appendix[A] 
\appendixtitle{Review of information thermodynamics}

We re-derive the second law of information thermodynamics and the conservation law of information flows, both of which are used in this study. The derivation is mainly based on \citet{Seifert12RepProgPhys}, \citet{Horowitz+Esposito14PhysRevX}, and \citet{Loos+Klapp20NewJPhys}. We show the derivation in detail because information thermodynamics is rarely applied to atmospheric and oceanic sciences, and comprehensive reviews for continuous-state systems are not readily available. Indeed, recent textbooks primarily focus on discrete-state Markov jump processes \citep{Peliti+Pigolotti21Book, Shiraishi23Book}, which are not directly applicable to the continuous-state climate system. Moreover, \citet{Ito16Book} explains the concept of information thermodynamics clearly despite its concise description. Although not straightforward, according to \citet{Ito16SciRep}, the theory in \citet{Ito16Book} is equivalent to the theory used in our study \citep{Horowitz+Esposito14PhysRevX, Loos+Klapp20NewJPhys}.

\subsection{Multivariate nonlinear SDEs} \label{subsec:app-N-SDEs}

We derive the second law of information thermodynamics for $N$-variable nonlinear stochastic differential equations \cite[SDEs, e.g., ][]{Gardiner09Book}.
\begin{align}
    {\rm d} x_i = - a_i(\mathbf{x}) {\rm d} t+ \sqrt{2D_i} \, {\rm d} W_i, \quad (i =1, \ldots, N), \label{eq:N-SDE}
\end{align}
where $\mathbf{x} := (x_1, \ldots, x_N)^{\rm T}$, $a_i(\mathbf{x})$ is a nonlinear function of $\mathbf{x}$, $D_i$ is a positive constant, ${\rm d} W_i$ is the differential of mutually independent Wiener processes for each $i$. Equation (\ref{eq:N-SDE}) assumes that the coefficient matrix $\bm{\mathsf{D}}$ for the Wiener processes is diagonal: $\bm{\mathsf{D}} = {\rm diag}(\sqrt{2D_1}, \ldots, \sqrt{2D_N})$. This assumption corresponds to the bipartite condition \citep{Horowitz+Esposito14PhysRevX} and is essential for the following derivation. For example, this condition is necessary for the transformation from Eq. (\ref{eq:def-flux-J}) to (\ref{eq:grad-p-by-flux-J}), and Eq. (\ref{eq:grad-p-by-flux-J}) is indispensable for the final inequality (\ref{eq:inequality-by-flux}). If $\bm{\mathsf{D}}$ is non-diagonal, a noise ${\rm d} W_i$ may change not only $x_i$ but also other state variables. By taking the matrix $\bm{\mathsf{D}}$ to be diagonal, it is guaranteed that ${\rm d} W_i$ contributes only to the fluctuation of $x_i$. Moreover, if $\bm{\mathsf{D}}$ is diagonal, each component can be taken positive without loss of generality. We expressed it as $\sqrt{2 D_i}$ to simplify later equations.

The probability distribution $p_{\rm tot}(\mathbf{x}, t)$ of the ensemble of solutions to these SDEs follows the Fokker-Planck equation \cite[FPE, e.g., ][]{Gardiner09Book}. The subscript ``$\rm tot$'' emphasizes that $p_{\rm tot}$ is the joint probability of all variables:
\begin{equation}
    \frac{\partial p_{\rm tot}(\mathbf{x},t)}{\partial t} =\sum_{i=1}^N \frac{\partial}{\partial x_i}\left[a_i(\mathbf{x}) p_{\rm tot}(\mathbf{x},t)\right]+ \sum_{i=1}^N D_{i} \frac{\partial^2}{\partial x_i^2} p_{\rm tot}(\mathbf{x},t) = - \sum_{i=1}^N \frac{\partial}{\partial x_i} J_i(\mathbf{x},t).
\end{equation}
In the rightmost side, the probability flux $\mathbf{J}$ is introduced, explicitly expressing the conservation of probability. The $i$-th component of $\mathbf{J}$ is defined as follows:
\begin{align}
    J_i(\mathbf{x},t) &:=-a_i(\mathbf{x}) p_{\rm tot}(\mathbf{x},t) - D_i \frac{\partial}{\partial x_i} p_{\rm tot}(\mathbf{x},t). \label{eq:def-flux-J}
\end{align}
Using $J_i$, the derivative $\partial p_{\rm tot}/\partial x_i$ is expressed as:
\begin{align}
    \frac{\partial p_{\rm tot}}{\partial x_i} &= -\frac{J_i}{D_i}-\frac{p_{\rm tot}\,a_i}{D_i}. \label{eq:grad-p-by-flux-J}
\end{align}
For boundary conditions for the FPE, we assume $p_{\rm tot}(\mathbf{x},t) = 0$ and $\mathbf{J}(\mathbf{x},t) = \mathbf{0}$ at sufficiently far away from the origin.

\subsection{Second law of information thermodynamics for subsystems}

We derive the second law of information thermodynamics by transforming the time derivative of the Shannon entropy $S_\zeta$ for a subsystem $z$ ($:= x_{\zeta}$) \citep{Horowitz+Esposito14PhysRevX,Loos+Klapp20NewJPhys}:
\begin{equation}
    S_\zeta := -\int {\rm d} z \; p_\zeta(z, t) \ln p_\zeta(z, t), \label{eq:deff-shannon-partial-z}
\end{equation}
where $\zeta$ represents any index from $1$ to $N$. We introduced $\zeta$ to distinguish it from other subscripts, such as $i$ and $j$, and to emphasize that $\zeta$ corresponds to $z (= x_\zeta)$. The marginal probability $p_\zeta$ is defined by the integral of the joint probability $p_{\rm tot}$:
\begin{equation}    
    p_\zeta(z, t) :=\int \left(\prod_{i=1, i\ne \zeta}^{N} {\rm d} x_i\right) \; p_{\rm tot}(\mathbf{x}, t),
\end{equation}
The time evolution of this marginal probability $p_\zeta$ is given by the integral of the convergence of the probability flux:
\begin{align}
    \frac{\partial}{\partial t} p_\zeta(z,t) &= \frac{\partial}{\partial t} \int \left(\prod_{i=1, i \ne \zeta}^{N} {\rm d} x_i \right)\; p_{\rm tot}(\mathbf{x}, t), \label{eq:dp_zeta-dt1} \\
    & =-\int  \left( \prod_{i=1, i \ne \zeta}^{N} {\rm d} x_i \right) \; \sum_{j=1}^N \frac{\partial}{\partial x_j} J_j(\mathbf{x}, t), \label{eq:dp_zeta-dt2} \\
    & =-\int \left(\prod_{i=1, i \ne \zeta}^{N} {\rm d} x_i \right)\; \frac{\partial}{\partial z} J_\zeta(\mathbf{x}, t). \label{eq:dp_zeta-dt3}
\end{align}
In the last equation, we integrate the derivative $\partial J_i / \partial x_i$ ($i \ne \zeta$) and use the boundary conditions, leading it to be zero. The point of the following derivation is to use the FPE via the joint probability $p_{\rm tot}(\mathbf{x},t)$ to deal with the marginal probability $p_\zeta(z,t)$.

We transform the time derivative of the Shannon entropy:
\begin{align}
    \frac{d}{d t} S_\zeta 
    & =-\int \frac{\partial p_\zeta}{\partial t} \ln p_\zeta \; {\rm d} z-\underbrace{\int \frac{\partial p_\zeta}{\partial t} {\rm d} z}_{=0}, \label{eq:trans-shannon-zeta1} \\
    & =\int {\rm d}z \int \left(\prod_{i=1, i \ne \zeta}^{N} {\rm d} x_i\right) \; \left[\frac{\partial}{\partial z} J_\zeta(\mathbf{x}, t)\right] \ln p_\zeta(z,t), \label{eq:trans-shannon-zeta2} \\
    & =\int  \underbrace{\left(\prod_{i=1}^{N} {\rm d} x_i\right)}_{= {\rm d}\mathbf{x}} \; \left[\frac{\partial J_\zeta(\mathbf{x}, t)}{\partial z}\right]\left[\ln \frac{p_\zeta(z,t)}{p_{\rm tot}(\mathbf{x}, t)}+\ln p_{\rm tot}(\mathbf{x}, t)\right], \label{eq:trans-shannon-zeta3} \\
    &= \dot{I}_{\zeta \leftarrow} + \int {\rm d}\mathbf{x} \; \frac{\partial J_\zeta(\mathbf{x}, t)}{\partial z} \ln p_{\rm tot}(\mathbf{x}, t). \label{eq:trans-shannon-zeta4}
\end{align}
In Eq. (\ref{eq:trans-shannon-zeta1}), the conservation of probability was used. In Eq. (\ref{eq:trans-shannon-zeta3}), the definition of $z$ is used: $z:=x_\zeta$. We define the information flow $\dot{I}_{i \leftarrow}$ as follows \citep{Allahverdyan+09JStatMech, Loos+Klapp20NewJPhys}:
\begin{equation}
  \dot{I}_{i \leftarrow} := \int {\rm d}\mathbf{x} \; \frac{\partial J_i(\mathbf{x}, t)}{\partial x_i} \ln \frac{p_i(x_i, t)}{p_{\rm tot}(\mathbf{x}, t)}, \label{eq:def-information-flow}
\end{equation}
where the marginal distribution for $x_i$ is
\begin{equation}
    p_i(x_i, t) :=\int \left(\prod_{j=1, j\ne i}^{N} {\rm d} x_j \right) \; p_{\rm tot}(\mathbf{x}, t). \label{eq:def-marginal-prob}
\end{equation}
The information flow $\dot{I}_{i \leftarrow}$ represents changes in the mutual information by the subsystem $x_i$. We further discuss information flow in Subsection \ref{subsec:information-flow-conservation} in Appendix A. In Eq. (\ref{eq:trans-shannon-zeta4}), the information flow $\dot{I}_{\zeta \leftarrow}$ (i.e., $i = \zeta$) was substituted.

We further transform the second term in Eq. (\ref{eq:trans-shannon-zeta4}):
\begin{align}
    \int {\rm d} \mathbf{x} \; \frac{\partial J_\zeta}{\partial z} \ln p_{\rm tot} & = -\int {\rm d} \mathbf{x} \; J_\zeta \frac{\partial\left(\ln p_{\rm tot}\right)}{\partial z}+ \int \left(\prod_{i=1,i\ne \zeta}^{N} {\rm d} x_i\right) \; \underbrace{\left[ J_\zeta \ln p_{\rm tot}\right]_{z=-\infty}^{z=+\infty}}_{=0}, \label{eq:trans-shannon-zeta-second1}\\
    & =-\int {\rm d} \mathbf{x} \; \frac{J_\zeta}{p_{\rm tot}} \frac{\partial p_{\rm tot}}{\partial z}, \label{eq:trans-shannon-zeta-second2}\\
    & =\left\langle-\frac{1}{p_{\rm tot}} \frac{\partial p_{\rm tot}}{\partial z} \circ \frac{{\rm d} z}{{\rm d} t}\right\rangle, \label{eq:trans-shannon-zeta-second3}\\
    & =\left\langle\frac{1}{p_{\rm tot}}\left(\frac{J_\zeta}{D_\zeta}+\frac{p_{\rm tot}\,a_\zeta}{D_\zeta}\right) \circ \frac{{\rm d} z}{{\rm d} t}\right\rangle, \label{eq:trans-shannon-zeta-second4}\\
    & =\left\langle\frac{J_\zeta}{p_{\rm tot} D_\zeta} \circ \frac{{\rm d} z}{{\rm d} t}\right\rangle+\left\langle\frac{a_\zeta}{D_\zeta} \circ \frac{{\rm d} z}{{\rm d} t}\right\rangle, \label{eq:trans-shannon-zeta-second5}
\end{align}
where $\circ$ represents the Stratonovich product \citep[e.g.,][]{Gardiner09Book} and Eq. (\ref{eq:grad-p-by-flux-J}) with $i=\zeta$ is substituted in Eq. (\ref{eq:trans-shannon-zeta-second3}) to replace $\partial p_{\rm tot} / \partial z$. We further explain the Stratonovich product and the conversion formula between $J_\zeta$ and ${{\rm d} z}/{{\rm d} t}$ used in transforming Eq. (\ref{eq:trans-shannon-zeta-second2}) to (\ref{eq:trans-shannon-zeta-second3}).

First, the Stratonovich product is defined as follows \citep[e.g.,][]{Gardiner09Book}:
\begin{equation}
    f(\mathbf{x}(t)) \circ {\rm d} W := \lim_{\Delta t \rightarrow 0} \frac{f(\mathbf{x}(t+\Delta t)) + f(\mathbf{x}(t))}{2} \left(W_{t+\Delta t} - W_t\right), \label{eq:def-stratonovich-2}
\end{equation}
where $f$ represents any smooth function. Using the Stratonovich product, the chain rules hold mathematically \citep{Gardiner09Book}, and the energy conservation law (i.e., the first law of thermodynamics) holds physically \citep{Sekimoto10Book}. The introduction of the Stratonovich product is also discussed in \citet{Shiraishi23Book}. Because of these properties, the Stratonovich product is commonly used in information thermodynamics. The Stratonovich product is necessary only for terms where the differential ${\rm d}W$ appears. For example, using Eq. (\ref{eq:def-stratonovich-2}), the product of $f(\mathbf{x}(t))$ and ${{\rm d} z}/{{\rm d} t}$ becomes:
\begin{align}
    f(\mathbf{x}(t)) \circ \frac{{\rm d} z}{{\rm d} t} &= f(\mathbf{x}(t)) \circ \left( -a_\zeta  + \sqrt{2 D_\zeta} \frac{{\rm d}W_\zeta}{{\rm d}t} \right), \\
    &= -f(\mathbf{x}(t))a_\zeta + \sqrt{2 D_\zeta} f(\bm{x}(t)) \circ \frac{{\rm d}W_\zeta}{{\rm d}t},
\end{align}
where ${\rm d} W_\zeta / {\rm d} t$ cannot be defined mathematically because of the non-smoothness of $W_\zeta(t)$ \citep{Gardiner09Book}, but in physics, ${\rm d} W_\zeta / {\rm d} t$ is often used for computational convenience as white Gaussian noise.

In the transformation from Eq. (\ref{eq:trans-shannon-zeta-second2}) to (\ref{eq:trans-shannon-zeta-second3}), the following formula was used \citep{Seifert12RepProgPhys}:
\begin{equation}
    \left\langle f(\mathbf{x}(t)) \circ \frac{{\rm d} x_i}{{\rm d} t} \right\rangle = \int {\rm d}\mathbf{x} \; f(\mathbf{x}) J_i(\mathbf{x},t). \label{eq:dot-flux}
\end{equation}
The statistical average of the rate of change of the state variable ${{\rm d} x_i}/{{\rm d} t}$ becomes the integral of the $i$-th component of the probability flux $J_i(\mathbf{x},t)$ in the phase space. To show this formula, we use the definition of the Stratonovich product [Eq. (\ref{eq:def-stratonovich-2})] and perform a Taylor expansion. To simplify the derivation, we ignore terms of order smaller than ${\rm d}t$.
\begin{align}
    {\rm d}t \left\langle f(\mathbf{x}(t)) \circ \frac{{\rm d} x_i}{{\rm d} t} \right\rangle
    &= \left\langle f(\mathbf{x}(t)) \circ {\rm d}x_i \right\rangle, \label{eq:trans-dot-flux1}\\
    &= \left\langle\left(f+\sum_{k=1}^N \frac{1}{2}\frac{\partial f}{\partial x_k} {\rm d} x_k\right) {\rm d} x_i\right\rangle, \label{eq:trans-dot-flux2}\\
    & =\left\langle f(x) {\rm d} x_i+\frac{1}{2}  \frac{\partial f}{\partial x_i}\left({\rm d}x_i\right)^2\right\rangle, \label{eq:trans-dot-flux3}\\
    &= {\rm d}t \left\langle -a_i\,f + D_i \frac{\partial f}{\partial x_i}\right\rangle, \label{eq:trans-dot-flux4}\\
    &= {\rm d}t \int {\rm d}\mathbf{x} \; J_i(\mathbf{x},t) f(\mathbf{x}). \label{eq:trans-dot-flux5}
\end{align}
From Eq. (\ref{eq:trans-dot-flux1}) to (\ref{eq:trans-dot-flux2}), the definition of the Stratonovich product was used, and a Taylor expansion was performed. From Eq. (\ref{eq:trans-dot-flux2}) to (\ref{eq:trans-dot-flux4}), the Ito rule was used, where $\langle {\rm d} x_k {\rm d} x_i\rangle = 0$ for $k \ne i$ due to the diagonal matrix $\bm{\mathsf{D}}$ in the SDEs [Eq. (\ref{eq:N-SDE})]. In Eq. (\ref{eq:trans-dot-flux5}), after integration by parts, the definition of the probability flux [Eq. (\ref{eq:def-flux-J})] was substituted. Finally, Eq. (\ref{eq:dot-flux}) is derived by comparing Eqs. (\ref{eq:trans-dot-flux1}) and (\ref{eq:trans-dot-flux5}).

To rewrite the numerator of the second term in Eq. (\ref{eq:trans-shannon-zeta-second5}), we define the following quantity:
\begin{align}
    \dot{Q}_i &:= -\left\langle a_i(\mathbf{x}) \circ \frac{{\rm d} x_i}{{\rm d} t} \right\rangle, \label{eq:def-Q} \\
    &=  \left\langle \left(\frac{{\rm d} x_i}{{\rm d} t} - \xi_i\right) \circ \frac{{\rm d} x_i}{{\rm d} t} \right\rangle, \label{eq:def-Q-white-noise}      
\end{align}
where $\xi_i := \sqrt{2 D_i} {\rm d} W_i / {\rm d} t$ and the SDE for $i$ [Eq. (\ref{eq:N-SDE})] is used to obtain the second line. The quantity $\dot{Q}_i$ can be regarded as the amount of heat dissipated by the subsystem $x_i$ to the surrounding environment per unit time \citep{Sekimoto10Book,Loos+Klapp20NewJPhys}.

Combining Eqs. (\ref{eq:trans-shannon-zeta4}), (\ref{eq:trans-shannon-zeta-second5}), (\ref{eq:dot-flux}), and (\ref{eq:def-Q}), we obtain
\begin{align}
    \underbrace{\frac{{\rm d} S_\zeta}{{\rm d} t}+\frac{\dot{Q}_\zeta}{D_\zeta}}_{=: \dot{\sigma}_\zeta}-\dot{I}_{\zeta \leftarrow} & =\left\langle\frac{J_\zeta}{p_{{\rm tot}} D_\zeta} \circ \frac{{\rm d} z}{{\rm d} t}\right\rangle =\int {\rm d} \mathbf{x} \; \frac{J_\zeta^2}{p_{{\rm tot}} D_\zeta} \ge 0,  \label{eq:inequality-by-flux}
\end{align}
where the equality holds when $J_\zeta = 0$, that is, in the case of the detailed balance being satisfied for $z$ \citep{Loos+Klapp20NewJPhys, Gardiner09Book}. Since the subscript $\zeta$ is arbitrary, the following inequality is derived:
\begin{equation}
    \dot{\sigma}_i = \frac{{\rm d} S_i}{{\rm d} t}+\frac{\dot{Q}_i}{D_i} \ge \dot{I}_{i \leftarrow}. \label{eq:second-law-partial}
\end{equation}
We defined the production rate of total entropy $\dot{\sigma}_i$ as the sum of the rate of change of the Shannon entropy ${\rm d} S_i / {\rm d} t$ of the subsystem $x_i$ and the rate of change of the heat $\dot{Q}_i/{D_i}$ dissipated by the subsystem $x_i$ to the surroundings. The production rate of total entropy $\dot{\sigma}_i$ is lower bounded by the information flow $\dot{I}_{i \leftarrow}$. If $\dot{I}_{i \leftarrow}$ is negative, $\dot{\sigma}_i$ can be negative, that is, the total entropy can decrease. In the usual thermodynamic setup, $D_i$ is equal to the temperature of the environment \citep{Sekimoto10Book,Loos+Klapp20NewJPhys}. Thus, Eq. (\ref{eq:second-law-partial}) is considered a generalization of the Clausius inequality. The equality holds when $J_i = 0$. In this case, the detailed balance is satisfied mathematically \citep{Gardiner09Book}, and the subsystem $x_i$ becomes an equilibrium state physically \citep{Loos+Klapp20NewJPhys}.

In the steady state, the time derivative of the Shannon entropy becomes zero, and the following holds:
\begin{equation}
    \dot{\sigma}_i = \frac{\dot{Q}_i}{D_i} \ge \dot{I}_{i \leftarrow}. \label{eq:second-law-partial-steady}
\end{equation}
The steady state is defined as a state in which joint probabilities of state variables between any time points depend only on the time differences (i.e., lags) and not on the absolute time itself. As a result, in the steady state, the marginal probability $p_i(x_i, t)$ does not change over time; thus, the time derivative of the Shannon entropy $S_i$ becomes zero.

\subsection{Conservation law of information flows} \label{subsec:information-flow-conservation}

We derive the conservation law of information flows \citep{Allahverdyan+09JStatMech,Loos+Klapp20NewJPhys}. We first define the mutual information $M$ \citep[e.g.,][]{Cover+Thomas05Book}:
\begin{equation}
    M := \int {\rm d} \mathbf{x} \; p_{\rm {tot }}(\mathbf{x}, t) \ln \left[\frac{p_{\rm {tot }}(\mathbf{x}, t)}{\prod_{i=1}^{N} p_i(x_i, t)} \right] \label{eq:def-mutual-information-N}
\end{equation}
Mutual information represents the magnitude of nonlinear correlations between subsystems. The value of mutual information is non-negative, and it becomes zero only when the subsystems are independent of each other.

We transform the time derivative of mutual information.
\begin{align}
    \frac{{\rm d} M}{{\rm d} t} &=\int {\rm d} \mathbf{x} \; \frac{\partial p_{\rm tot}}{\partial t} \ln \left(\frac{p_{\rm tot}}{\prod_{i=1}^{N} p_i}\right) +\int {\rm d} \mathbf{x} \; p_{\rm tot}\left(\frac{\partial}{\partial t} \ln p_{\rm tot} -\sum_{i=1}^N \frac{\partial}{\partial t} \ln p_i\right), \label{eq:trans-mutual-info1}\\
    & =\int {\rm d} \mathbf{x} \; \frac{\partial p_{\rm tot}}{\partial t} \ln \left(\frac{p_{\rm tot}}{\prod_{i=1}^{N} p_i}\right) +\underbrace{\int {\rm d} \mathbf{x} \; \frac{\partial p_{\rm tot}}{\partial t}}_{=0}-\sum_{i=1}^N \underbrace{\int {\rm d} x_i \; \frac{\partial p_i}{\partial t}}_{=0}, \label{eq:trans-mutual-info2} \\
    &=-\int {\rm d} \mathbf{x}\; \left(\sum_{j=1}^N \frac{\partial J_j(\mathbf{x},t)}{\partial x_j}\right) \ln \left[ \frac{p_{\rm tot}(\mathbf{x},t)}{\prod_{i=1}^{N} p_i(x_i,t)}\right], \label{eq:trans-mutual-info3} \\
    &= \sum_{j=1}^N \left[\int {\rm d} \mathbf{x} \; \frac{\partial J_j}{\partial x_j} \left(\sum_{i=1}^N \ln p_i\right)-\int {\rm d} \mathbf{x} \; \frac{\partial J_j}{\partial x_j} \ln p_{\rm tot} \right], \label{eq:trans-mutual-info4}\\
    &= \sum_{j=1}^N\left[\int {\rm d} \mathbf{x} \; \frac{\partial J_j}{\partial x_j} \ln p_j -\int {\rm d} \mathbf{x} \; \frac{\partial J_j}{\partial x_j} \ln p_{\rm tot}\right], \label{eq:trans-mutual-info5}
\end{align}
In Eq. (\ref{eq:trans-mutual-info2}), the conservation of probability and the definition of marginal probability were used. From Eq. (\ref{eq:trans-mutual-info4}) to (\ref{eq:trans-mutual-info5}), for all $i$ such that $i \ne j$, the integral of $J_j$ is executable and the result becomes zero due to the boundary conditions. Consequently, only the terms with $i=j$ remain in Eq. (\ref{eq:trans-mutual-info5}).

The conservation law of information flows is obtained by substituting Eq. (\ref{eq:def-information-flow}) into (\ref{eq:trans-mutual-info5}):
\begin{equation}
    \frac{{\rm d} M}{{\rm d} t} = \sum_{i=1}^{N} \dot{I}_{i\leftarrow}. \label{eq:conservation-information}
\end{equation}
The sum of changes in the mutual information by the subsystem $x_i$ (i.e., the information flow $\dot{I}_{i \leftarrow}$) over all $i$ becomes the temporal change of the mutual information of the entire system. The information flow $\dot{I}_{i \leftarrow}$ is interpreted as the sum of the inflow and outflow of information from all the other subsystems to $x_i$.

In the steady state, since the joint probability $p_{\rm tot}$ does not change over time, the time derivative of mutual information becomes zero. As a result, the following holds:
\begin{equation}
    0 = \sum_{i=1}^{N} \dot{I}_{i \leftarrow}. \label{eq:conservation-information-steady}
\end{equation}

The source of information flow is uniquely determined when the entire system consists of only two variables \citep{Allahverdyan+09JStatMech}. Here, the two random variables are denoted by uppercase letters $X_t$ and $Y_t$, and their realizations are denoted by lowercase letters $x_t$ and $y_t$, respectively, where the time dependence is explicitly denoted by the subscript $t$. In this case, the mutual information [Eq. (\ref{eq:def-mutual-information-N})] is expressed as follows:
\begin{equation}
    M(X_t,Y_t) := \int {\rm d}x_t{\rm d}y_t \; p(x_t, y_t) \ln \left[\frac{p(x_t,y_t)}{p(x_t)p(y_t)} \right]. \label{eq:def-mutual-information-2}
\end{equation}
The dependence on the variables $X_t$ and $Y_t$ is explicitly written inside the parentheses as $M(X_t,Y_t)$, where $M(X_t,Y_t)$ is symmetric with respect to $X_t$ and $Y_t$. In the bivariate case, the information flows are expressed as follows \citep{Allahverdyan+09JStatMech}:
\begin{align}
\dot{I}_{X \leftarrow Y} &= \lim_{\Delta t \rightarrow 0}\frac{M(X_{t+\Delta t},Y_t) - M(X_t,Y_t)}{\Delta t}, \label{eq:def-information-flow-Y-to-X}\\
\dot{I}_{Y \leftarrow X} &= \lim_{\Delta t \rightarrow 0}\frac{M(X_t,Y_{t+\Delta t}) - M(X_t,Y_t)}{\Delta t}. \label{eq:def-information-flow-X-to-Y}
\end{align}
Based on the Leibniz rule for differentiation, the sum of these information flows satisfies the conservation law (\ref{eq:conservation-information}). The variation of mutual information due to the time change of the random variable $X$ (or $Y$) is defined as the information flow to $X$ (or $Y$). Equation (\ref{eq:def-information-flow-Y-to-X}) can be transformed into $\dot{I}_{X \leftarrow}$ defined by Eq. (\ref{eq:def-information-flow}), and Eq. (\ref{eq:def-information-flow-X-to-Y}) is also transformed into $\dot{I}_{Y \leftarrow}$ in the same manner \citep{Allahverdyan+09JStatMech}. From the conservation law of information flows in the steady state, namely $\dot{I}_{X \leftarrow Y} + \dot{I}_{Y \leftarrow X} = 0$, we see that information flows can be negative or positive.

\subsection{Second law of thermodynamics for the entire system} \label{subsec:second-law-total-system}

We derive the second law of thermodynamics for the entire system and confirm that focusing on subsystems is essential in information thermodynamics \citep{Sagawa+Ueda13NewJPhys}. As a preparation, we define the Shannon entropy of the entire system:
\begin{equation}
    S_{\rm tot} :=-\int {\rm d} \mathbf{x} \; p_{\rm tot}(\mathbf{x},t) \ln p_{\rm tot}(\mathbf{x},t). \label{eq:def-shannon-entropy-total}
\end{equation}

In general, additivity does not hold for Shannon entropy. The degree to which additivity does not hold is evaluated by mutual information.
\begin{align}
    \left(\sum_{i=1}^N S_i\right)- M &= -\left[\sum_{i=1}^N \int {\rm d} x_i \; p_i(x_i,t) \ln p_i(x_i,t)\right]-\int {\rm d}\mathbf{x} \; p_{\rm tot}(\mathbf{x},t) \ln \left[\frac{p_{\rm tot}(\mathbf{x},t)}{\prod_{i=1}^N p_i(x_i,t)}\right],\label{eq:trans-additivity-entropy-N1}\\
    & =-\left(\sum_{i=1}^N\int {\rm d} \mathbf{x} \; p_{\rm tot} \ln p_i \right)-\int {\rm d} \mathbf{x} p_{\rm tot}  \ln \left( \frac{p_{\rm tot}}{\prod_{i=1}^{N} p_i} \right), \label{eq:trans-additivity-entropy-N2}\\
    & =-\int {\rm d} \mathbf{x} \; p_{\rm tot} \ln p_{\rm tot},\label{eq:trans-additivity-entropy-N3}\\
    & = S_{\rm tot}, \label{eq:trans-additivity-entropy-N4}
\end{align}
where we use the definitions of $S_i$, $M$, and $S_{\rm tot}$ in Eqs. (\ref{eq:deff-shannon-partial-z}), (\ref{eq:def-mutual-information-N}), and (\ref{eq:def-shannon-entropy-total}), respectively. From Eq. (\ref{eq:trans-additivity-entropy-N1}) to (\ref{eq:trans-additivity-entropy-N2}), the definition of marginal probability [Eq. (\ref{eq:def-marginal-prob})] was applied to the first term. The Shannon entropies of the subsystems are additive only when the mutual information $M$ is zero.

We derive the second law for the entire system. Since Eq. (\ref{eq:second-law-partial}) holds for any $i$, it can be summed over all $i$. Applying Eqs. (\ref{eq:conservation-information}) and (\ref{eq:trans-additivity-entropy-N4}), we obtain the following:
\begin{align}
    \dot{\sigma}_{\rm tot} := \frac{{\rm d} S_{\rm tot}}{{\rm d} t}+\sum_{i=1}^{N} \frac{\dot{Q}_i}{D_i} \ge 0. \label{eq:second-law-total}
\end{align}
The first term in the middle represents the rate of change of the Shannon entropy of the entire system, and the second term represents the sum of the rates of change of the heat dissipated to the environment. In Eq. (\ref{eq:second-law-total}), the production rate of total entropy for the entire system is lower bounded by zero, whereas in Eq. (\ref{eq:second-law-partial}), the production rate of total entropy for a subsystem is lower bounded by the information flow,  which can be negative. This information flow appears due to the non-additivity of Shannon entropy \citep{Sagawa+Ueda13NewJPhys}. Historically, the second law of thermodynamics [Eq. (\ref{eq:second-law-total})] was derived first \citep{Seifert05PRL}, and the second law of information thermodynamics [Eq. (\ref{eq:second-law-partial})] was derived later \citep{Horowitz+Esposito14PhysRevX,Loos+Klapp20NewJPhys}.

\subsection{Interpretation of heat} \label{subsec:interpretation-heat}

The quantity $\dot{Q}_i$ in Eq. (\ref{eq:def-Q}) is regarded as the heat dissipated by a subsystem $x_i$ to the environment per unit time \citep{Sekimoto10Book}. When the environment is always in a thermal equilibrium state, this heat is equal to the entropy production rate of the environment \citep{Peliti+Pigolotti21Book, Shiraishi23Book}. However, this interpretation is not always possible.

Generally, in systems described by stochastic differential equations, noise is not always due to thermal fluctuations. For example, in the dynamical system considered in the current research [Eqs. (\ref{eq:dynamical-sys-TG}) and (\ref{eq:dynamical-sys-TK})], noise represents the influence of small-scale disturbances in the atmosphere and ocean (Section \ref{sec:dynamical-system}\ref{subsec:valid-dynamical-system}). Moreover, it is unclear whether such small-scale disturbance fields are always in an equilibrium state, although this point may be investigated using equilibrium statistical mechanics for geophysical fluids \citep{Majda+Wang06Book, Bouchet+Venaille12PhysRep, Campa+14Book}. Indeed, the meanders of the Gulf Stream and the Kuroshio Current are consistent with analysis using equilibrium statistical mechanics \citep{Venaille+Bouchet11JPO}, which implies that small-scale disturbance fields in both current regions can also be analyzed using equilibrium theory. Currently, however, such an analysis is not sufficient, and the interpretation of $\dot{Q}_i$ needs to be considered for each application. In particular, $\dot{\sigma}_i = \dot{Q}_i / D_i$ holds in the steady state [Eq. (\ref{eq:second-law-partial-steady})], making this interpretation essential.

The present study has analyzed the dynamical system [Eqs. (\ref{eq:dynamical-sys-TG}) and (\ref{eq:dynamical-sys-TK})] using the second law of information thermodynamics as a mathematical tool (Section \ref{sec:analysis}). Although the second law is always derived for SDEs as shown above, mathematically deriving the second law is different from obtaining a meaningful physical insight from it. We have regarded $\dot{\sigma}_G$ ($= \dot{Q}_G / D_G$) and $\dot{\sigma}_K$ ($= \dot{Q}_K / D_K$) as indicators of the contribution of interactions to the temporal tendencies of the SSTs. This interpretation does not rely on the interpretation of $\dot{Q}_i$ as heat. As a similar example, \citet{Ito+Sagawa15NatComm} considered $\dot{Q}_i$ as the robustness of signal transmission in a cell (i.e., not as heat) and analyzed the processing efficiency of cell signaling using information thermodynamics. The application of information thermodynamics in this manner might yield interesting results for other atmospheric and oceanic systems. Investigating further applications is an important topic for future research in atmospheric and oceanic sciences.

\appendix[B]
\appendixtitle{Theoretical formulas used in Sections \ref{sec:dynamical-system} and \ref{sec:analysis}}

\subsection{Correspondence between linear dynamical system and AR1 model} \label{subsec:linear-SDE-and-AR1}

We convert the AR1 model to the linear dynamical system by comparing the means and covariances. The obtained theoretical formulas were used in Section \ref{sec:dynamical-system}\ref{subsec:regression-analysis}.

The linear dynamical system is described in the vector form as follows \citep{Gardiner09Book}:
\begin{equation}
    {\rm d} \mathbf{x} = -\bm{\mathsf{A}} \, \mathbf{x} \, {\rm d}t + \bm{\mathsf{B}} \, {\rm d} \mathbf{W}, \label{eq:N-SDE-vec-form}
\end{equation}
where $\mathbf{x}$ is an $N$-dimensional real vector variable, time $t$ is continuous, and $\mathbf{W}$ represents $N$ Wiener processes that are independent of each other. We consider here a general case of $N \ge 1$, although the case of $N=2$ was used in Section \ref{sec:dynamical-system}. The constant real matrix $\bm{\mathsf{A}}$ is assumed to have only eigenvalues with positive real parts, and $\bm{\mathsf{B}}$ is a constant real matrix. Under these conditions, $\mathbf{x}$ has a stationary solution \citep{Gardiner09Book}. Integrating Eq. (\ref{eq:N-SDE-vec-form}) over a time step $\Delta t$, the following solution is obtained \citep{Gardiner09Book}:
\begin{equation}
    \mathbf{x}(t+\Delta t)=e^{-\bm{\mathsf{A}} \Delta t} \mathbf{x}(t)+\int_0^{\Delta t} e^{-\bm{\mathsf{A}}(\Delta t-s)} \bm{\mathsf{B}} \; {\rm d} \mathbf{W}(s).
\end{equation}

The multivariate AR1 model is given as follow \citep{Hamilton94Book}:
\begin{equation}
    \mathbf{y}_{t+\Delta t} = \bm{\mathsf{C}} \, \mathbf{y}_{t} + \mathbf{\varepsilon}_t,
\end{equation}
where $\mathbf{y}$ is an $N$-dimensional real vector variable and $\mathbf{\varepsilon}$ represents vector white noise. The absolute values of all eigenvalues of the real matrix $\bm{\mathsf{C}}$ are assumed to be smaller than $1$, which is the condition for stationarity \citep{Hamilton94Book}. Time $t$ is discrete and takes steps at a constant interval $\Delta t$. The covariance matrix $\bm{\mathsf{V}}$ of the vector white noise is generally non-diagonal and is expressed as
\begin{equation}
  \langle \bm{\varepsilon}_t \bm{\varepsilon}_s^{\rm T}\rangle = \begin{cases}
      \bm{\mathsf{V}} & (t = s), \\
      0 & (\text{otherwise}),
  \end{cases}
\end{equation}
where $\langle \cdot \rangle$ represents the statistical average over noise.

We obtain the condition for the mean values of $\mathbf{x}({t+\Delta t})$ and $\mathbf{y}_{t+\Delta t}$ being equal:
\begin{equation}
  \bm{\mathsf{C}} = e^{-\bm{\mathsf{A}} \Delta t}, \label{eq:cond-equal-mean-sde-ar1}
\end{equation}
where $\mathbf{x}(t) = \mathbf{y}_t$ is assumed. We also obtain the condition for the covariances of $\mathbf{x}({t+\Delta t})$ and $\mathbf{y}_{t+\Delta t}$ being equal:
\begin{equation}
    \bm{\mathsf{V}} = \int_0^{\Delta t} {\rm d} s\; e^{ -\bm{\mathsf{A}} (\Delta t-s) } \bm{\mathsf{B}} \bm{\mathsf{B}}^{\rm T} e^{-\bm{\mathsf{A}}^{\rm T}(\Delta t-s)}. \label{eq:cond-equal-cov-sde-ar1}
\end{equation}
Under stationarity, Eqs. (\ref{eq:cond-equal-mean-sde-ar1}) and (\ref{eq:cond-equal-cov-sde-ar1}) become independent of $t$.

The regression analysis with the AR1 model determines the matrices $\bm{\mathsf{C}}$ and $\bm{\mathsf{V}}$, where the matrix $\bm{\mathsf{V}}$ is estimated from the residuals of regression. Using Eqs. (\ref{eq:cond-equal-mean-sde-ar1}) and (\ref{eq:cond-equal-cov-sde-ar1}), $\bm{\mathsf{A}}$ and $\bm{\mathsf{B}}$ are then obtained. This method requires that the AR1 model and the dynamical system share the same statistical properties over the time interval $\Delta t$. Since the linear SDE [Eq. (\ref{eq:N-SDE-vec-form})] describes a Gaussian process, its parameters are uniquely determined from the mean and covariance. Table \ref{table:estimated-coeffs} shows the coefficients of the dynamical system [Eqs. (\ref{eq:dynamical-sys-TG}) and (\ref{eq:dynamical-sys-TK})], which were obtained by applying Eqs. (\ref{eq:cond-equal-mean-sde-ar1}) and (\ref{eq:cond-equal-cov-sde-ar1}) to the regression result with AR1.

\subsection{Theoretical expressions of lag correlation coefficients} \label{subsec:theory-lag-corr}

We show here the theoretical expressions of lag correlation coefficients used in Sections \ref{sec:dynamical-system}\ref{subsec:reproduction-bcs} and \ref{sec:analysis}\ref{subsec:regime-diagrams}. The dynamical system [Eqs. (\ref{eq:dynamical-sys-TG}) and (\ref{eq:dynamical-sys-TK})] describes a Gaussian process, so the statistical properties of the steady state are completely determined by the lag covariance matrix \citep{Gardiner09Book}. By normalizing the lag covariance, we obtain the lag correlation coefficient, which is an important indicator for the synchronization between the Gulf Stream and the Kuroshio Current \citep{Kohyama+21Science}.

We first show the instantaneous covariance matrix:
\begin{equation}    
\bm{\mathsf{\Sigma}} =
    \begin{pmatrix}
        \Sigma_{GG} & \Sigma_{GK} \\
        \Sigma_{KG} & \Sigma_{KK} \\
    \end{pmatrix}
    =
    \begin{pmatrix}
        \langle T_G T_G\rangle & \langle T_G T_K\rangle \\
        \langle T_K T_G\rangle & \langle T_K T_K\rangle \\
    \end{pmatrix}, \label{eq:def-sigma-cov-mat}
\end{equation}
where the off-diagonal components are equal, $\Sigma_{GK} = \Sigma_{KG}$. Each component is given as follows:
\begin{align}
    \Sigma_{GG} &= \frac{(r_K^2+r_G r_K-c_{G \leftarrow K} c_{K \leftarrow G})D_G + c_{G \leftarrow K}^2 D_K}{(r_G+r_K)(r_G r_K-c_{G \leftarrow K} c_{K \leftarrow G})},\label{eq:sigma-gg} \\
    \Sigma_{KK} &= \frac{(r_G^2+r_G r_K-c_{G \leftarrow K} c_{K \leftarrow G})D_K + c_{K \leftarrow G}^2 D_G}{(r_G+r_K)(r_G r_K-c_{G \leftarrow K} c_{K \leftarrow G})},\label{eq:sigma-kk} \\
    \Sigma_{GK} &= \Sigma_{KG} = \frac{r_K c_{K \leftarrow G} D_G + r_G c_{G \leftarrow K} D_K}{(r_G+r_K)(r_G r_K-c_{G \leftarrow K} c_{K \leftarrow G})}. \label{eq:sigma-gk}
\end{align}
Details of the derivation for the instantaneous covariance can be found in \citet{Gardiner09Book}, directly leading to Eqs. (\ref{eq:sigma-gg})--(\ref{eq:sigma-gk}).

Using the instantaneous covariance $\bm{\mathsf{\Sigma}}$, the lag covariance matrix is given as follows \citep{Gardiner09Book}:
\begin{equation}
    \langle T_i(t_1) T_j(t_2) \rangle = \begin{cases}
        \left( e^{-\bm{\mathsf{A}}(t_1-t_2)} \bm{\mathsf{\Sigma}} \right)_{ij} & (t_1 \ge t_2), \\
        \left( \bm{\mathsf{\Sigma}} e^{-\bm{\mathsf{A}}^{\rm T}(t_2-t_1)} \right)_{ij} & (t_1 < t_2),
    \end{cases} \label{eq:lag-cov}
\end{equation}
where
\begin{equation}
    \bm{\mathsf{A}}= \left(
      \begin{matrix}
          r_G & -c_{G\leftarrow K} \\
          -c_{K\leftarrow G} & r_K \\
      \end{matrix}
    \right).
\end{equation}
The indices $G$ and $K$ are subscripts for the Gulf Stream and the Kuroshio Current, respectively. In Section  \ref{sec:dynamical-system}\ref{subsec:reproduction-bcs}, Fig. \ref{fig:lag-correlation} shows the lag correlation coefficients computed by normalizing the lag covariances with the instantaneous variances:
\begin{align}
    \rho(\Delta t) &= \frac{\left( e^{-\bm{\mathsf{A}}\Delta t} \bm{\mathsf{\Sigma}} \right)_{GK}}{\sqrt{\Sigma_{GG}\Sigma_{KK}}} \quad (\Delta t \ge 0), \\
    \rho(\Delta t) &= \frac{\left( \bm{\mathsf{\Sigma}} e^{-\bm{\mathsf{A}}^{\rm T}|\Delta t|} \right)_{GK}}{\sqrt{\Sigma_{GG}\Sigma_{KK}}} \quad (\Delta t < 0),
\end{align}
where $\Delta t$ represents the lag.

\subsection{Theoretical expressions of information thermodynamic quantities} \label{subsec:theory-entropy-info-flow}

This section derives the expressions for information thermodynamic quantities, such as $\dot{\sigma}_G$, used in Section \ref{sec:analysis}. In other words, the equations shown below are the specific forms of the theoretical equations in Appendix A applied to the dynamical system, Eqs. (\ref{eq:dynamical-sys-TG}) and (\ref{eq:dynamical-sys-TK}). Details of the derivation are discussed in \citet{Loos+Klapp20NewJPhys}.

As a preparation, we first examine the expressions for the variances and covariances. In the steady state, the probability distribution does not change in time, so the following formulas hold:
\begin{align}
    \frac{1}{2}\frac{{\rm d}}{{\rm d}t} \langle T_G T_G\rangle &= \left\langle T_G \circ \frac{{\rm d} T_G}{{\rm d} t}\right\rangle =0, \label{eq:dot-TG-squared}\\
    \frac{1}{2}\frac{{\rm d}}{{\rm d}t} \langle T_K T_K\rangle &= \left\langle T_K \circ \frac{{\rm d} T_K}{{\rm d} t}\right\rangle =0, \label{eq:dot-TK-squared}\\
    \frac{{\rm d}}{{\rm d}t} \langle T_G T_K \rangle &= \left\langle T_G \circ \frac{{\rm d} T_K}{{\rm d} t}\right\rangle + \left\langle T_K \circ \frac{{\rm d} T_G}{{\rm d} t}\right\rangle =0. \label{eq:dot-TG-TK}
\end{align}
The temporal derivatives ${\rm d} T_G / {\rm d} t$ or ${\rm d} T_K/ {\rm d} t$ include white Gaussian noise $\xi_G$ or $\xi_K$, respectively. Thus, the Stratonovich product $\circ$ needs to be used (Appendix A). The symbol $\circ$ is omitted in Eqs. (\ref{eq:sigma_G})--(\ref{eq:correlation-TG-TK-in-steady-state}), (\ref{eq:sigma_G_decomposition}), and (\ref{eq:sigma_K_decomposition}) because the Stratonovich product is not explained in Section \ref{sec:analysis}. By substituting the governing equations (\ref{eq:dynamical-sys-TG}) and (\ref{eq:dynamical-sys-TK}), each term in Eq. (\ref{eq:dot-TG-TK}) is given by
\begin{align}
  \left\langle T_G \circ \frac{{\rm d} T_K}{{\rm d} t} \right\rangle &= -r_K \langle T_K T_G\rangle + c_{K \leftarrow G} \langle T_G T_G \rangle, \label{eq:TG-dot-TK}\\
  \left\langle T_K \circ \frac{{\rm d} T_G}{{\rm d} t} \right\rangle &= -r_G \langle T_G T_K\rangle + c_{G \leftarrow K} \langle T_K T_K \rangle,\label{eq:TK-dot-TG}
\end{align}
where the variances and covariances on the right-hand sides are given by Eqs. (\ref{eq:sigma-gg})--(\ref{eq:sigma-gk}).

We next provide the expressions for the entropy production rates. From Eqs. (\ref{eq:dynamical-sys-TG}), (\ref{eq:dynamical-sys-TK}), (\ref{eq:def-Q}), and (\ref{eq:second-law-partial-steady}), the following equations are derived for the steady state:
\begin{align}
  \dot{\sigma}_G = \frac{\dot{Q}_G}{D_G} &= -\frac{r_G}{D_G} \left\langle T_G \circ \frac{{\rm d} T_G}{{\rm d} t}\right\rangle + \frac{c_{G\leftarrow K}}{D_G} \left\langle T_K \circ \frac{{\rm d} T_G}{{\rm d} t}\right\rangle, \label{eq:specific-form-sigma-G1} \\
  &= \frac{c_{G\leftarrow K}}{D_G} \left\langle T_K \circ \frac{{\rm d} T_G}{{\rm d} t}\right\rangle, \label{eq:specific-form-sigma-G2} \\
  \dot{\sigma}_K = \frac{\dot{Q}_K}{D_K} &= -\frac{r_K}{D_K} \left\langle T_K \circ \frac{{\rm d} T_K}{{\rm d} t}\right\rangle + \frac{c_{K\leftarrow G}}{D_K} \left\langle T_G \circ \frac{{\rm d} T_K}{{\rm d} t}\right\rangle, \label{eq:specific-form-sigma-K1} \\
  &= \frac{c_{K\leftarrow G}}{D_K} \left\langle T_G \circ \frac{{\rm d} T_K}{{\rm d} t}\right\rangle, \label{eq:specific-form-sigma-K2} 
\end{align}
where Eqs. (\ref{eq:dot-TG-squared}) and (\ref{eq:dot-TK-squared}) are substituted. Equations (\ref{eq:specific-form-sigma-G2}) and (\ref{eq:specific-form-sigma-K2}) are the same as Eqs. (\ref{eq:sigma_G}) and (\ref{eq:sigma_K}), respectively. Equation (\ref{eq:relation-sigmaG-sigmaK}) is obtained using Eqs. (\ref{eq:dot-TG-TK}), (\ref{eq:specific-form-sigma-G2}), and (\ref{eq:specific-form-sigma-K2}).

The expressions for the information flows are easily derived in the steady state. Using the property that the probability distribution does not change in time, the time derivative of the Shannon entropy is zero. Thus, from Eq. (\ref{eq:trans-shannon-zeta4}), the following holds in general:
\begin{align}
  \dot{I}_{i \leftarrow} &= - \int {\rm d}\mathbf{x} \; \frac{\partial J_i}{\partial x_i} \ln p_{\rm tot}, \label{eq:info-flow-steady-simple1}\\
    &= \int {\rm d}\mathbf{x} \; J_i \frac{\partial \ln p_{\rm tot}}{\partial x_i} + \int \left(\prod_{j=1, j\ne i}^N {\rm d}x_j\right) \; \underbrace{\left[ J_i p_{\rm tot}\right]_{x_i=-\infty}^{x_i=\infty}}_{=0}, \label{eq:info-flow-steady-simple2}\\
    &= \left\langle \frac{\partial \ln p_{\rm tot}}{\partial x_i} \circ \frac{{\rm d} x_i}{{\rm d} t} \right\rangle, \label{eq:info-flow-steady-simple3}
\end{align}
where the boundary conditions and Eq. (\ref{eq:dot-flux}) are used here. We only need to evaluate the derivatives of $\ln p_{\rm tot}$ in Eq. (\ref{eq:info-flow-steady-simple3}). For a linear SDE, the stationary distribution is a Gaussian \citep{Gardiner09Book}, and the derivative of its logarithm can be expressed using the variances and covariances. The information flows are then derived as
\begin{align}
  \dot{I}_{G \leftarrow K} &= -(\bm{\mathsf{\Sigma}}^{-1})_{KG} \left\langle T_K \circ \frac{{\rm d} T_G}{{\rm d} t} \right\rangle,\\
  \dot{I}_{K \leftarrow G} &= -(\bm{\mathsf{\Sigma}}^{-1})_{GK} \left\langle T_G \circ \frac{{\rm d} T_K}{{\rm d} t} \right\rangle,
\end{align}
where Eqs. (\ref{eq:dot-TG-squared}) and (\ref{eq:dot-TK-squared}) are substituted, and the inverse of the covariance matrix $\bm{\mathsf{\Sigma}}$ is denoted as $\bm{\mathsf{\Sigma}}^{-1}$. The off-diagonal components of $\bm{\mathsf{\Sigma}}^{-1}$ are $(\bm{\mathsf{\Sigma}}^{-1})_{KG}$ and $(\bm{\mathsf{\Sigma}}^{-1})_{GK}$, which are equal because $\bm{\mathsf{\Sigma}}$ is symmetric. Using the specific expression of this off-diagonal component, we obtain Eqs. (\ref{eq:I_G}) and (\ref{eq:I_K}).

The information flows and entropy production rates become zero when $c_{G \leftarrow K}D_K = c_{K \leftarrow G}D_G$ \citep{Loos+Klapp20NewJPhys}. This fact can be directly verified by substituting Eqs. (\ref{eq:sigma-gg})--(\ref{eq:sigma-gk}), (\ref{eq:TG-dot-TK}), and (\ref{eq:TK-dot-TG}) into Eqs. (\ref{eq:sigma_G})--(\ref{eq:I_K}). A simpler method to show this is through the application of the second law of information thermodynamics (see Appendix D). In the case of $c_{G \leftarrow K}D_K = c_{K \leftarrow G}D_G$, the two subsystems, namely the Gulf Stream and the Kuroshio Current, become symmetric. Mathematically, the detailed balance is satisfied and the probability flux $\mathbf{J}$ is zero \citep{Loos+Klapp20NewJPhys}.

\appendix[C]
\appendixtitle{Histograms of estimated quantities by the moving block bootstrap method}

We quantified the uncertainty of the estimated coefficients using the moving block bootstrap method \citep{Mudelsee14Book} in Section \ref{sec:dynamical-system}\ref{subsec:regression-analysis}, where the number of resampled time series was set to $2000$. We discuss here the histograms of the entropy production rates and information flows obtained from the coefficients estimated by the bootstrap method. 

\begin{figure}[t]
    \centering
    \noindent\includegraphics[width=16.45cm]{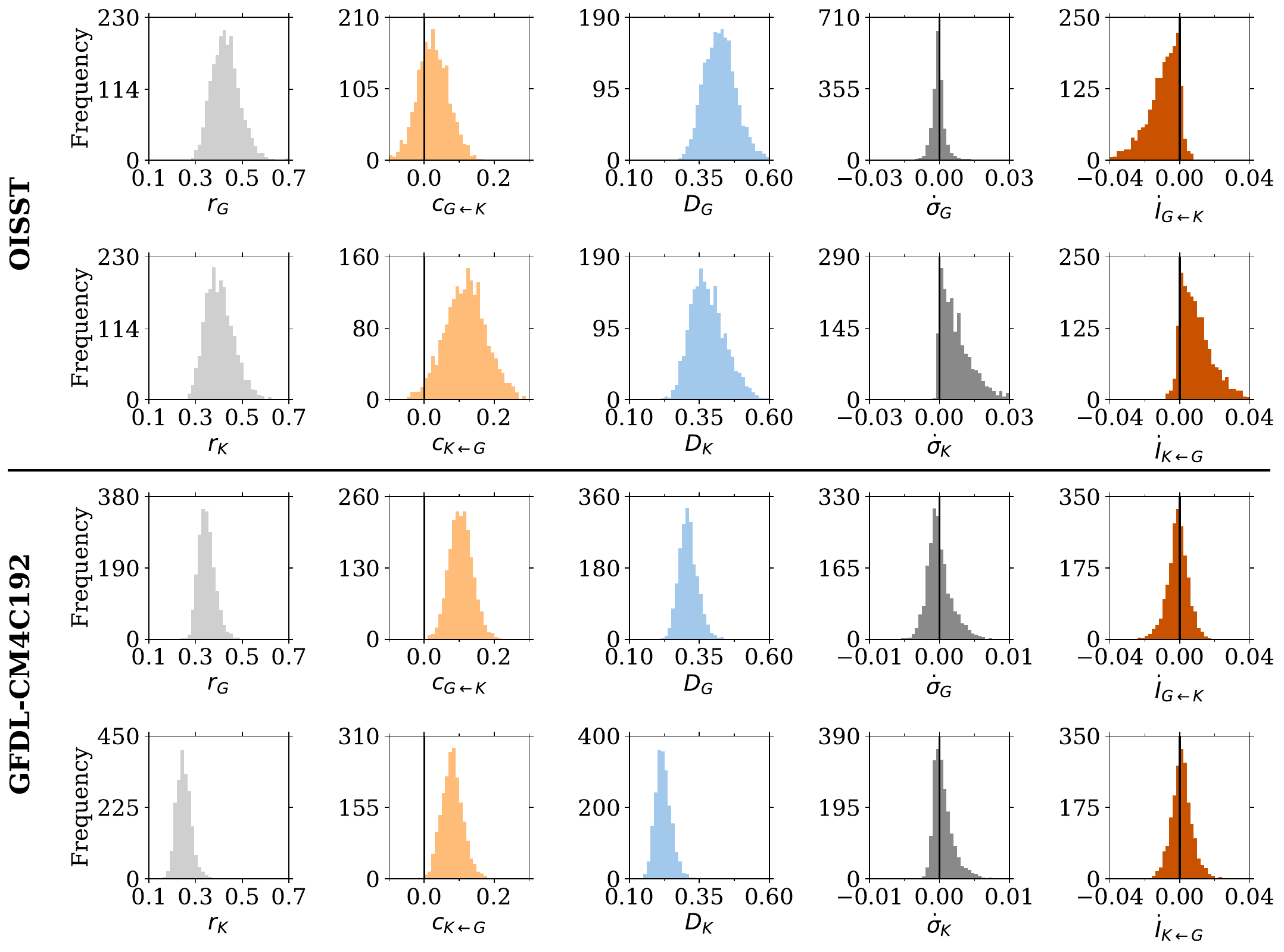}
    \caption{Histograms of all estimated quantities obtained from the moving block bootstrap method with $2000$ resampled time series for the OISST and GFDL-CM4C192 data. The quantities include all coefficients of the dynamical system, the entropy production rates, and the information flows. The vertical lines are at zero.}
    \label{fig:all-histos-estimates}
\end{figure}

Figure \ref{fig:all-histos-estimates} shows the histograms of all estimated quantities, including the coefficients of the dynamical system, where the number of samples in each histogram is $2000$. The 95\% confidence intervals in Table \ref{table:estimated-coeffs} were obtained from these histograms. We focus on the entropy production rates and information flows, which are not listed in Table \ref{table:estimated-coeffs}. The conservation of information flows, $\dot{I}_{G \leftarrow K} + \dot{I}_{K \leftarrow G} = 0$ [Eq. (\ref{eq:conservation-info-flow-TG-TK})], always holds. That is, for each resampled time series, $\dot{I}_{K \leftarrow G} = - \dot{I}_{G \leftarrow K}$. The histograms suggest that $\dot{I}_{G \leftarrow K} < 0$ and $\dot{I}_{K \leftarrow G} > 0$ for the OISST data, whereas these information flows take both signs for the GFDL-CM4C192 data. These results are also found in Figs. \ref{fig:regime-diagram-OISST}b, \ref{fig:regime-diagram-OISST}e, \ref{fig:regime-diagram-GFDL}b, and \ref{fig:regime-diagram-GFDL}e.

We next discuss the entropy production rates. Equation (\ref{eq:relation-sigmaG-sigmaK}) always holds. Thus, when $c_{G \leftarrow K} > 0$ and $c_{K \leftarrow G} > 0$, $\dot{\sigma}_K$ and $\dot{\sigma}_G$ have opposite signs to each other. For the GFDL-CM4C192 data, this opposite-sign relationship was confirmed in $1996$ out of $2000$ samples. On the other hand, for the OISST data, this relationship was confirmed in $1413$ samples. The other $587$ samples are in the region where $c_{G \leftarrow K} < 0$ and $c_{K \leftarrow G} > 0$. In this region, both $\dot{\sigma}_G$ and $\dot{\sigma}_K$ are positive and have the same signs. This result is also found in Fig. \ref{fig:regime-diagram-OISST}, where the error bars cross the lines of $c_{G \leftarrow K}=0$.

In the region $c_{G \leftarrow K} < 0$, the interpretation of Maxwell's demon systems is not possible because of the same signs of $\dot{\sigma}_G$ and $\dot{\sigma}_K$, but an interesting phenomenon still occurs in terms of information thermodynamics, which is called feedback cooling \citep{Horowitz+Sandberg14NewJPhys, Loos+Klapp20NewJPhys}. For the interpretation of Maxwell's demon systems, the signs of the entropy production rates need to be opposite to each other (Sections \ref{sec:maxwell-demon}\ref{subsec:resolution-maxwell-demon} and \ref{sec:maxwell-demon}\ref{subsec:autonomous-maxwell-demon}). As for the Kuroshio Current, the entropy production rate and information flow are positive, as in the case of Maxwell's demon. Thus, even in the scenario of feedback cooling, there is no change in the interpretation for the Kuroshio Current. The difference arises in the Gulf Stream, which has a positive entropy production rate ($\dot{\sigma}_G >0$). The Gulf Stream still exhibits a negative information flow, which indicates that the Gulf Stream is interpreted as being controlled by the Kuroshio Current, like in Maxwell's demon. However, this control is done by negative feedback and reduces the effect of noise \citep{Horowitz+Sandberg14NewJPhys, Loos+Klapp20NewJPhys}. As a result, the SST magnitude of the Gulf Stream is reduced (Fig. \ref{fig:regime-diagram-OISST}c). This reduction is considered as being achieved by discarding heat to the environment, leading to the positive entropy production rate ($\dot{\sigma}_G >0$). Even in this parameter region, the correlation between $T_G$ and $T_K$ is positive (Fig. \ref{fig:regime-diagram-corr}), suggesting that the BCS occurs. However, the interpretation is slightly different. The synchronization is achieved by the Kuroshio Current, which reduces the SST fluctuations of the Gulf Stream caused by atmospheric and oceanic noise.

\appendix[D]
\appendixtitle{Constraint based on the second law of information thermodynamics}

The analysis in Section \ref{sec:analysis} is based only on the signs of entropy production rates and information flows, without explicitly using the inequalities between them (i.e., the second law of information thermodynamics). Here, we derive an inequality for the instantaneous correlation coefficient using the second law of information thermodynamics, Eqs. (\ref{eq:second-law-TG}) and (\ref{eq:second-law-TK}). The derived inequality is a constraint that the dynamical system [Eqs. (\ref{eq:dynamical-sys-TG}) and (\ref{eq:dynamical-sys-TK})] must satisfy.

The second law of information thermodynamics is expressed as the following two inequalities:
\begin{align}
    \frac{c_{G \leftarrow K}}{D_G}\left\langle T_K \circ \frac{{\rm d}T_G}{{\rm d}t}\right\rangle \geq \frac{\Sigma_{GK}}{|\bm{\mathsf{\Sigma}}|}\left\langle T_K \circ \frac{{\rm d}T_G}{{\rm d}t}\right\rangle, \label{eq:TG-second-law-corr-style}\\
    \frac{c_{K \leftarrow G}}{D_K}\left\langle T_G \circ \frac{{\rm d}T_K}{{\rm d}t}\right\rangle \geq \frac{\Sigma_{GK}}{|\bm{\mathsf{\Sigma}}|}\left\langle T_G \circ \frac{{\rm d}T_K}{{\rm d}t}\right\rangle. \label{eq:TK-second-law-corr-style}
\end{align}
Equation (\ref{eq:TG-second-law-corr-style}) is obtained by substituting Eqs. (\ref{eq:sigma_G}) and (\ref{eq:I_G}) into (\ref{eq:second-law-TG}); likewise, Eq. (\ref{eq:TK-second-law-corr-style}) is obtained by substituting Eqs. (\ref{eq:sigma_K}) and (\ref{eq:I_K}) into (\ref{eq:second-law-TK}). From the steady-state condition [Eq. (\ref{eq:dot-TG-TK})], either $\langle T_K \circ {\rm d}T_G/{\rm d}t \rangle$ or $\langle T_G \circ {\rm d}T_K/{\rm d}t \rangle$ is positive, while the other is negative. Here, we assume $\langle T_K \circ{\rm d}T_G/{\rm d}t \rangle < 0$. The above two inequalities are then combined into one:
\begin{equation}
    \frac{c_{G \leftarrow K}}{D_G} \le \frac{\Sigma_{G K}}{\Sigma_{K K} \Sigma_{G G}-\left(\Sigma_{G K}\right)^2} \le \frac{c_{K \leftarrow G}}{D_K},
\end{equation}
where we use $|\bm{\mathsf{\Sigma}}| = \Sigma_{K K} \Sigma_{G G}-\left(\Sigma_{G K}\right)^2$. Including the other case where $\langle T_K \circ {\rm d}T_G/{\rm d}t \rangle > 0$, the final inequality becomes
\begin{equation}
    \min \left\{ \frac{c_{K \leftarrow G}}{D_K}, \frac{c_{G \leftarrow K}}{D_G} \right\}\le \frac{\Sigma_{G K}}{\Sigma_{K K} \Sigma_{G G}-\left(\Sigma_{G K}\right)^2} \le \max \left\{ \frac{c_{K \leftarrow G}}{D_K}, \frac{c_{G \leftarrow K}}{D_G} \right\}.
\end{equation}
When $c_{K \leftarrow G}/{D_K}$ and ${c_{G \leftarrow K}}/{D_G}$ are equal, it corresponds to the diagonal lines in Figs \ref{fig:regime-diagram-OISST}, \ref{fig:regime-diagram-GFDL}, and \ref{fig:regime-diagram-corr}. Crossing these lines reverses the magnitude relation between $c_{K \leftarrow G}/{D_K}$ and ${c_{G \leftarrow K}}/{D_G}$. The signs of $\langle T_K \circ {\rm d}T_G/{\rm d}t \rangle$ and $\langle T_G \circ {\rm d}T_K/{\rm d}t \rangle$ are also reversed across these lines.

The obtained inequality is transformed into an inequality for the instantaneous correlation coefficient. We define the instantaneous correlation coefficient as follows:
\begin{equation}
    \rho_0 := \frac{\Sigma_{GK}}{\sqrt{\Sigma_{GG}\Sigma_{KK}}}.
\end{equation}
Assuming the time series are standardized, the variances $\Sigma_{GG}$ and $\Sigma_{KK}$ are equal to $1$. The following inequality is then obtained:
\begin{equation}
    \min \left\{ \frac{c_{K \leftarrow G}}{D_K}, \frac{c_{G \leftarrow K}}{D_G} \right\}\le \frac{\rho_0}{1-(\rho_0)^2} \le \max \left\{ \frac{c_{K \leftarrow G}}{D_K}, \frac{c_{G \leftarrow K}}{D_G} \right\}, \label{eq:inequality-corr}
\end{equation}
where the function $f(x) = x /(1-x^2)$ is monotonically increasing in the interval $(-1, 1)$, that is, $f(x)$ is invertible. The inequality (\ref{eq:inequality-corr}) means that $f(\rho_0)$ is bounded by $c_{K \leftarrow G}/{D_K}$ and ${c_{G \leftarrow K}}/{D_G}$, suggesting that $\rho_0$ increases as the interaction coefficients become larger or the noise amplitudes become smaller. 

The inequality (\ref{eq:inequality-corr}) is a condition that the correlation coefficient $\rho_0$ must satisfy; thus, this inequality can be regarded as a constraint on the dynamical system. It is possible to obtain the explicit form of $\rho_0$. However, this explicit form requires the use of the complex formulas (\ref{eq:sigma-gg}), (\ref{eq:sigma-kk}), and (\ref{eq:sigma-gk}). The second law of information thermodynamics provides the inequality (\ref{eq:inequality-corr}) without relying on these complex formulas. In other words, the inequality implies the dependence of $\rho_0$ on the parameters, such as $c_{G \leftarrow K}$ and $D_G$, without knowing the explicit form of $\rho_0$.

\bibliographystyle{ametsocV6}
\bibliography{references}

\end{document}